%% file: main.tex
\documentclass[%
 reprint,
 amsmath,amssymb,
 aps,
 pra,
]{revtex4-2}

\usepackage{graphicx,float}
\usepackage{dcolumn}
\usepackage{bm}
\usepackage{hyperref}

\usepackage{tikz}
\usetikzlibrary{positioning}
\usetikzlibrary{arrows.meta}
\usetikzlibrary{calc,decorations,decorations.markings,decorations.text}

\begin{document}

\title{Shape Morphing of Planar Liquid Crystal Elastomers}

\author{Daniel Castro}
\author{Hillel Aharoni}
\email{hillel.aharoni@weizmann.ac.il}
\affiliation{Department of Physics of Complex Systems, Weizmann Institute of Science, Rehovot 76100, Israel
}%
\date{\today}

\begin{abstract}
We consider planar liquid crystal elastomers: two dimensional objects made of anisotropic responsive materials, that upon activation remain flat however change their planar shape. We derive a closed form, analytical solution based on the implicit linearity featured by this subclass of deformations. Our solution provides the nematic director field on an arbitrary domain starting with two initial director curves.
We discuss the different gauges choices for this problem, and the inclusion of disclinations in the nematic order. Finally, we propose several applications and useful design principles based on this theoretical framework.
\end{abstract}


\maketitle

A self-shaping surface is a thin sheet, made of natural or artificial environmentally-responsive materials or metamaterials, that is designed to undergo a specific shape change upon an external actuation. Such objects have been thoroughly studied in recent years, both at the fundamental and at the applicative level. Among the systems studied are plant tissues \cite{dervaux-ben-amar,armon-efrati-kupferman}, Hydrogels \cite{Klein-Efrati-Sharon,Kim-Hanna}, smart textiles \cite{Hu}, self-folding origami \cite{Santangelo}, inflatables \cite{Siefert}, and many more.

One class of self-shaping materials that has been extensively studied in recent years is that of liquid crystal elastomers (LCEs) \cite{warner-book}. Such materials, when actuated, undergo a local shrinking/expansion along predetermined local principal directions at every point. While the magnitude of this deformation is constant throughout the entire material, the principal shrinking direction (the nematic director field) may vary throughout the sheet.
Determining the implicit geometry induced by a particular two-dimensional director field (also know as the forward problem) has been solved \cite{aharoni-sharon-kupferman, mostajeran1} and amply explored \cite{duffy2,modes-warner,mostajeran2}. Likewise, the inverse problem of determining the LCE director field that will deform into a desired geometry, has been shown \cite{griniasty-aharoni-efrati,deturck-yang, gevirtz93,aharoni2018universal} to be solvable locally in the form of a system of nonlinear hyperbolic partial differential equations (PDEs).

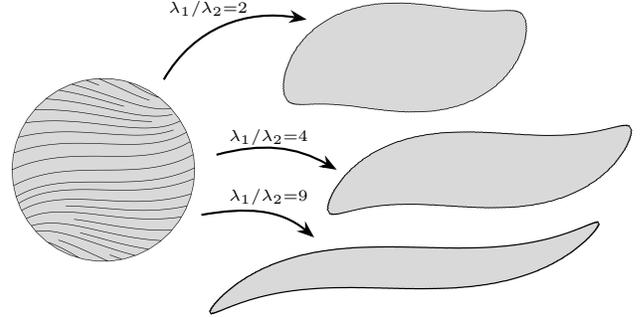
\begin{figure}
    \centering 
    \begin{tikzpicture}
        \node[inner sep=0pt] at (-4,0) {\includegraphics[width=.15\textwidth]{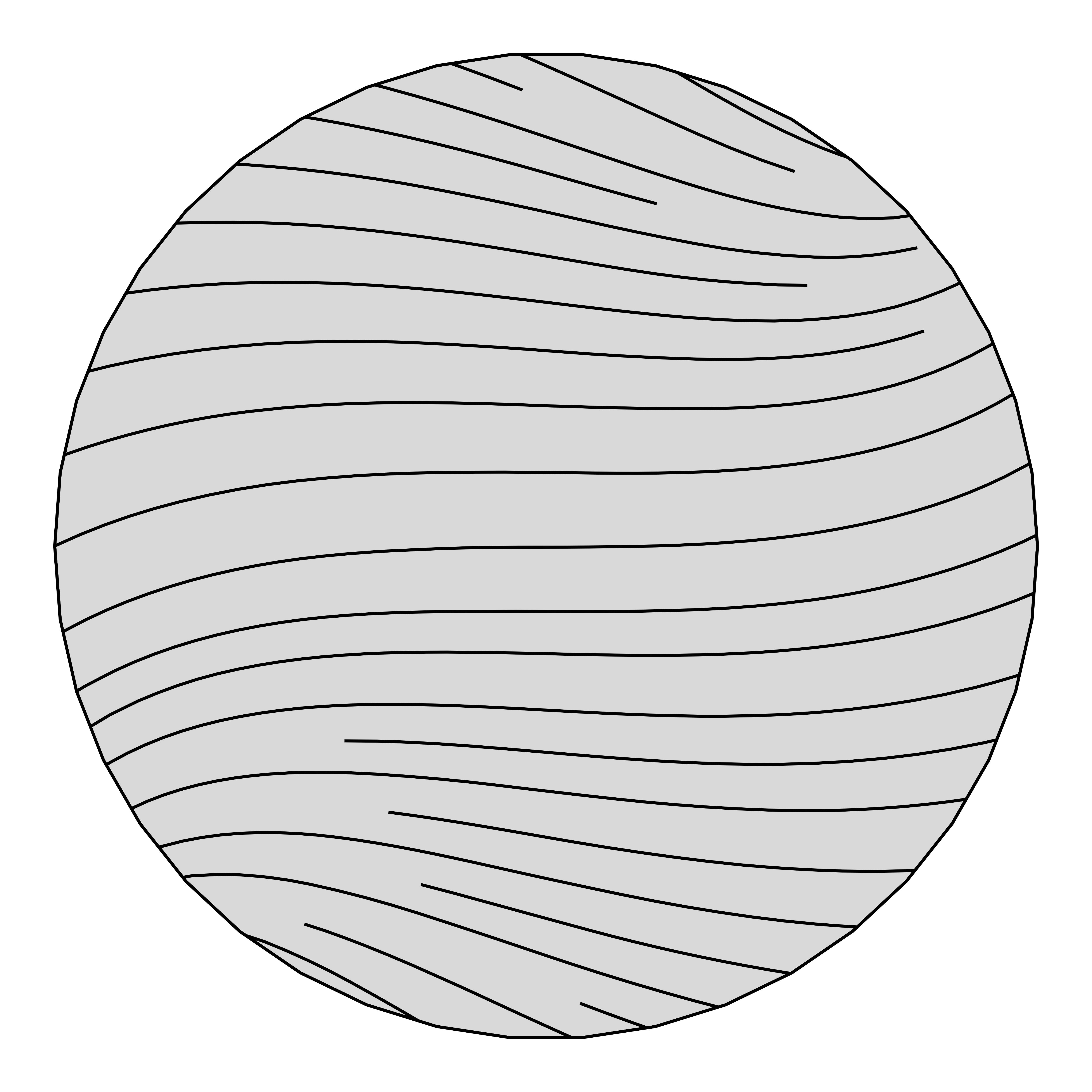}}; 
        \node[inner sep=0pt] at (0,1.5) {\includegraphics[width=.2\textwidth]{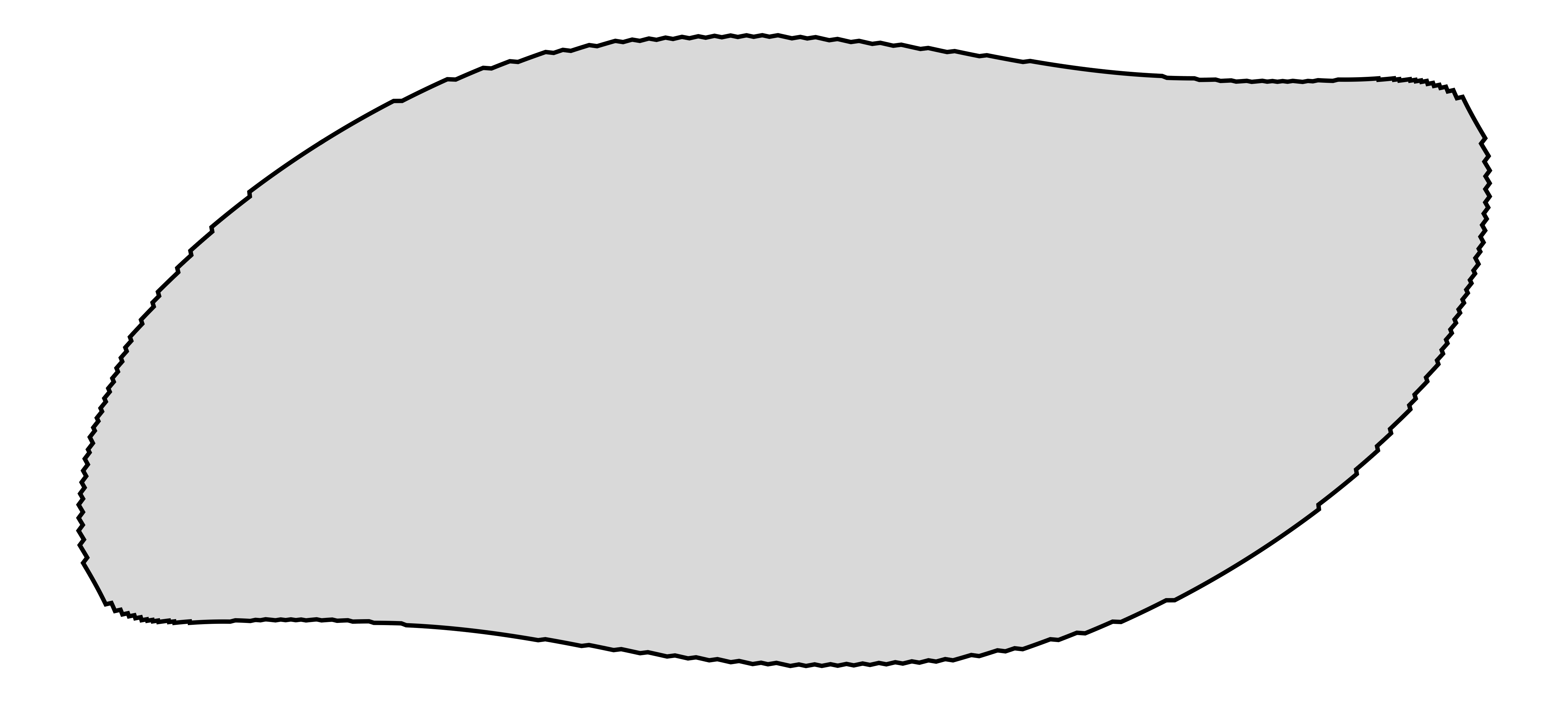}};
        \node[inner sep=0pt] at (1,-0.0) {\includegraphics[width=.25\textwidth]{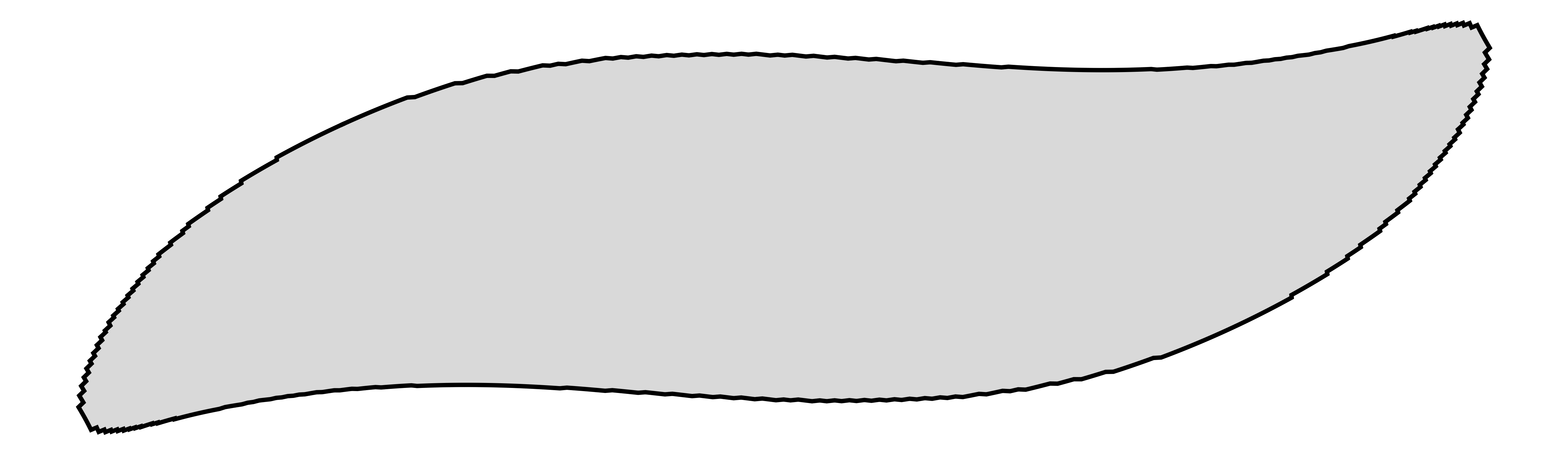}};
        \node[inner sep=0pt] at (0,-1.3)  {\includegraphics[width=.32\textwidth]{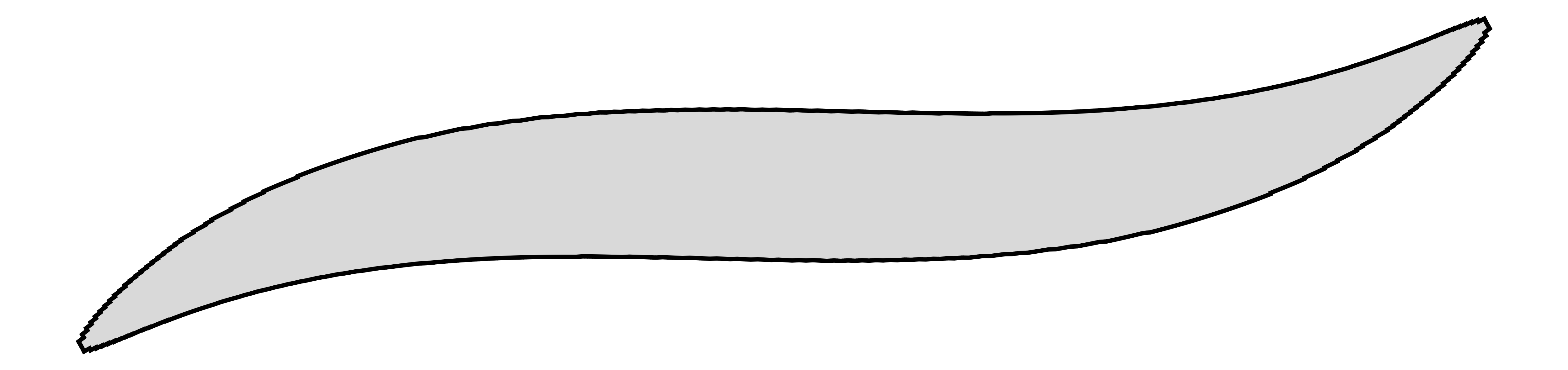}};
        \node[] at (-2.6,2.1) {$^{\lambda_1/\lambda_2=2}$}; \node[] at (-1.8,0.4) {$^{\lambda_1/\lambda_2=4}$}; \node[] at (-1.8,-0.4) {$^{\lambda_1/\lambda_2=9}$}; 
        \draw [-{Stealth[scale=1]},thick] (-3.2,1.2) [out=60,in=170] to (-1.3,2);      
        \draw [-{Stealth[scale=1]},thick] (-2.5,0.2) [out=15,in=150] to (-0.9,-0.0);
        \draw [-{Stealth[scale=1]},thick] (-2.7,-0.6) [out=10,in=140] to (-1.2,-0.9);   
    \end{tikzpicture}
    \caption{Planar liquid crystal elastomer (PLCE) deformations of a circular domain. The elongations along (by factor $\lambda_1$) and perpendicular ($\lambda_2$) to the director field deform the sheet and change the shape of its boundary, without buckling out of the plane (regardless of the sheet's thickness).}
    \label{fig:example}
\end{figure}

In this paper, our objects of interest are flat LCEs that \emph{remain flat} upon actuation, however their planar shape is deformed into a sequence of new shapes as a function of the actuation parameter, as exemplified in Fig.~\ref{fig:example}. We shall henceforth refer to these as planar LCEs, or PLCEs.
Even though the experimental realization of such sheets is not different from the case of generic, out-of-plane-deforming LCE sheets, the mathematical treatment is substantially simplified. We show that the absence of Gaussian curvature in the target geometry implies linearity of the PDEs governing the problem, therefore allowing for a closed, exact integral solution for the director field.

It's worth noting that such solutions, and the mappings from the plane to itself associated with them, are well known in the mathematical literature as constant principal strain (CPS) mappings, and significant results have been derived for them using analytical methods \cite{gevirtz92,gevirtz02,gevirtz01,gevirtz08}. Of cardinal importance to our context is Gevirtz's capability theorem \cite{gevirtz02}, which states that CPS mappings cannot transform a given domain into an arbitrary second one. The immediate conclusion is that some planar shape deformations are just not possible in LCEs, regardless of how extreme the local deformation gets. The inverse problem for planar domain deformations is not generically solvable. 

\paragraph*{The model-}
The director field imprinted on the initial surface is typically written as $\hat{\mathbf{n}}=\left(\cos\theta,\sin\theta \right)$ in Cartesian coordinates, and the induced Gaussian curvature of the actuated surface is written as some function of $\theta$ and its derivatives \cite{aharoni-sharon-kupferman, mostajeran1}. However, as emphasized in \cite{aharoni-sharon-kupferman}, except for particular, highly symmetric configurations, this set up is not convenient for the solution of the inverse problem, even in the flat cases that we are considering.  Alternatively, it is useful and in many aspects natural to use a coordinate system based on the integral curves of the director field and their perpendiculars \cite{niv-efrati, griniasty-aharoni-efrati}. In these coordinates, the director field is everywhere tangent to the local coordinate frame. Denoting by $u$ and $v$ the coordinates for the $\hat{\mathbf{n}}$ and $\hat{\mathbf{n}}_\perp$ curves, respectively, one may write the pre-actuated state as $\mathbf{r}\left( u,v\right)=\left( x\left( u,v\right),y\left( u,v\right)\right)$. The tangency condition is expressed as
\begin{align}
    \frac{\partial\mathbf{r}}{\partial u}=\alpha\hat{\mathbf{n}},\quad \frac{\partial\mathbf{r}}{\partial v}=\beta\hat{\mathbf{n}}_\perp,
    \label{eq:I:eq:embe}
\end{align} 
with some scale factors $\alpha\left(u,v\right)$ and  $\beta\left(u,v\right)$.

By construction, the length element of the preactuated sheet in these coordinates reads $\mathrm{d}s^2=\alpha^2\mathrm{d}u^2+\beta^2\mathrm{d}v^2$. The deformation upon actuation is a local contraction/expansion along the director field $\hat{\mathbf{n}}$ by a factor $\lambda_1$, and along the perpendicular $\hat{\mathbf{n}}_\perp$ by $\lambda_2$. This results in an actuated geometry of the exact form (in the same $uv$ coordinates), with uniformly rescaled scale factors $\alpha_\mathsf{A}=\lambda_1\alpha$ and $\beta_\mathsf{A}=\lambda_2\beta$. The compatibility conditions that impose zero Gaussian curvature in the initial sheet and $K_\mathsf{A}$ in the actuated one take the form \cite{griniasty-aharoni-efrati}
\begin{subequations}\label{eq:I:ab-sys} 
\begin{align}
\frac{1}{\beta}\frac{\partial b}{\partial v} & =b^{2}-\frac{K_\mathsf{A}}{\lambda_1^{-2}-\lambda_2^{-2}}\\
\frac{1}{\alpha}\frac{\partial s}{\partial u} & =-s^{2}-\frac{K_\mathsf{A}}{\lambda_1^{-2}-\lambda_2^{-2}},
\label{eq:I:ab1} 
\end{align}   
\end{subequations}
where $b$ and $s$ are the nematic bend and splay \cite{niv-efrati, griniasty-aharoni-efrati}:
\begin{equation}\label{eq:I:bs-def} 
b=\nabla\times\hat{\mathbf{n}}=-\frac{\partial_v\alpha}{\alpha\beta},\quad
s=\nabla\cdot\hat{\mathbf{n}}=\frac{\partial_u\beta}{\alpha\beta}.
\end{equation}
Replacing Eq.~(\ref{eq:I:bs-def}) into Eq.~(\ref{eq:I:ab-sys}), one obtains a self-contained PDE system for only $\alpha$ and $\beta$:
\begin{subequations}\label{eq:I:absy}
\begin{align}
\frac{\partial }{\partial v}\left(\frac{1}{\beta}\frac{\partial \alpha}{\partial v} \right) & =\alpha\beta\:\mathcal{K}\label{eq:I:ab11} \\
\frac{\partial }{\partial u}\left(\frac{1}{\alpha}\frac{\partial \beta }{\partial u} \right) & =-\alpha\beta\:\mathcal{K},\label{eq:I:ab2} 
\end{align}
\end{subequations}
with $\mathcal{K}=\left(\lambda_1^{-2}-\lambda_2^{-2} \right)^{-1}K_\mathsf{A}$.

As discussed in \cite{griniasty-aharoni-efrati}, these equations are a hyperbolic set whose characteristic curves are the $u,v$-lines themselves. Different types of initial conditions may be added to make a well-posed integrable problem. For our purposes, it is useful to set a Goursat problem \cite{griniasty-aharoni-efrati}; the initial value data is given along two intersecting characteristic curves. Namely, we consider a protocol in which we are given two plane curves that intersect perpendicularly. We then wish to design a PLCE such that one of the input curves is everywhere parallel and the other everywhere perpendicular to the nematic director. In our $uv$ coordinate system, these curves would correspond to $v=v_0$ and $u=u_0$. We are free to choose parametrization along these curves, namely $\alpha_0\left(u\right)\equiv\alpha\left(u,v_0\right)$ and $\beta_0\left(v\right)\equiv\beta\left(u_0,v\right)$, respectively (we later discuss this gauge freedom in detail). These initial conditions, together with Eq.~(\ref{eq:I:absy}), make a well-posed Goursat problem, to which a unique solution exists locally.

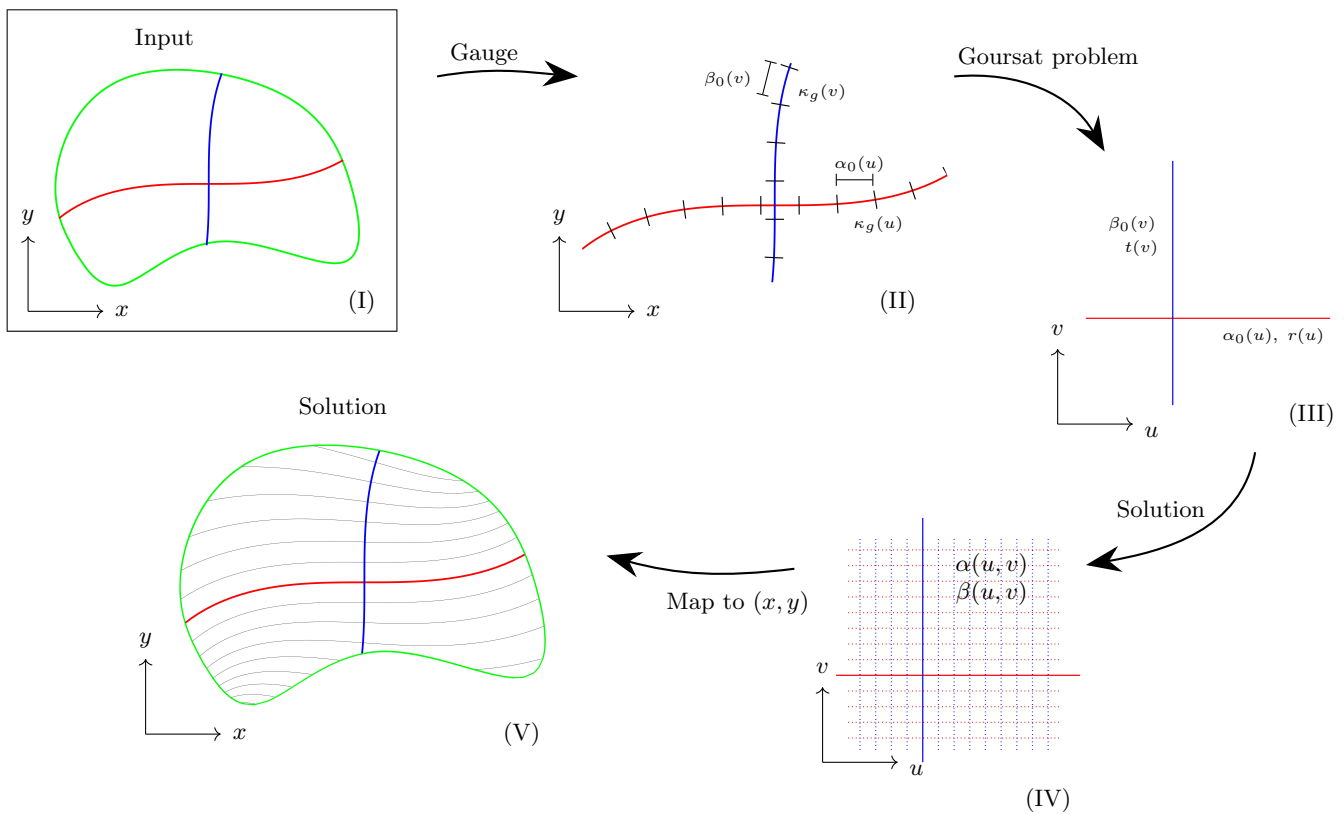
\begin{figure*}
   \centering 
    \begin{tikzpicture}[scale=2.5]  
        \node[inner sep=0pt] at (0.0,2.5) {\includegraphics[width=.3\textwidth]{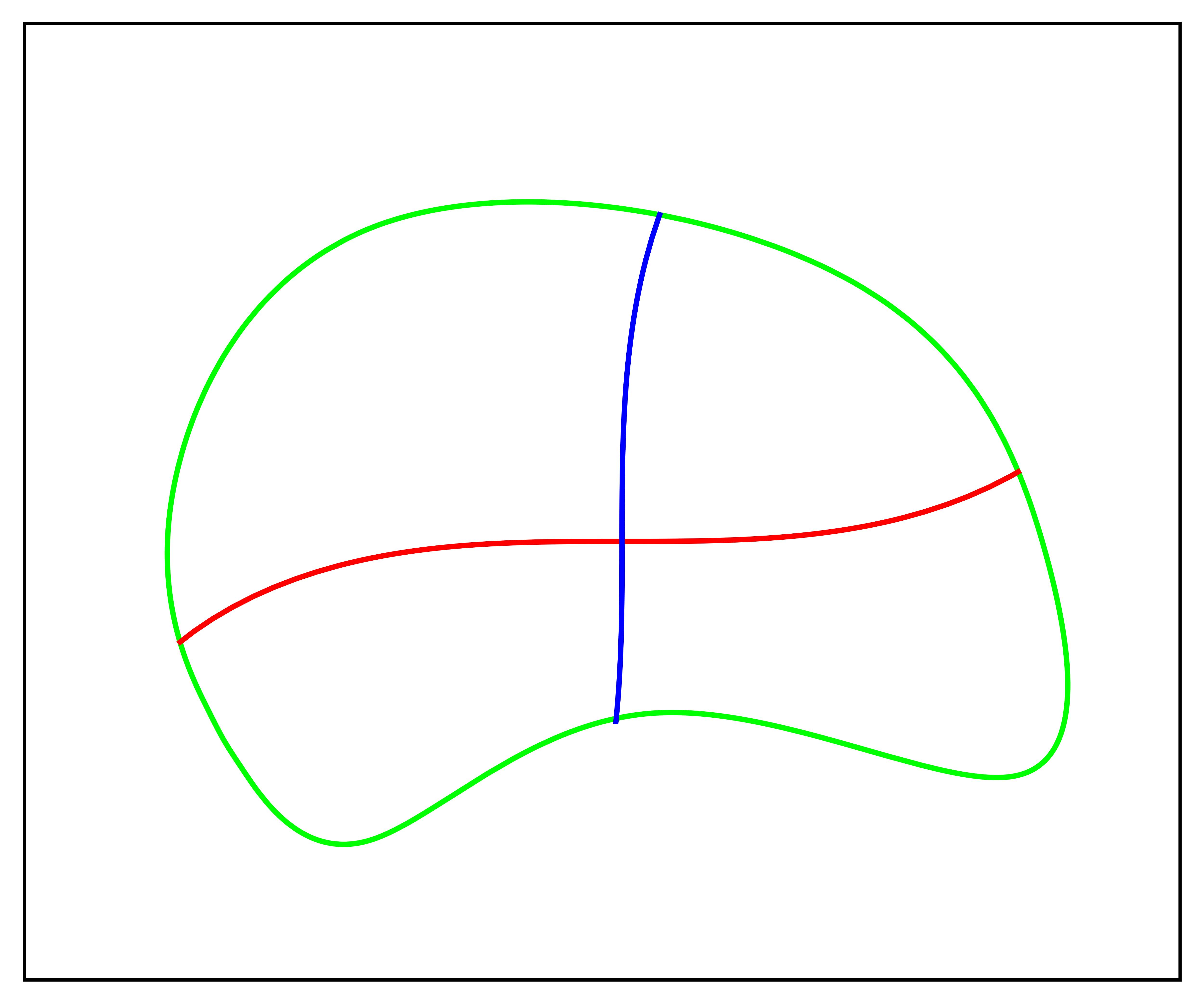}};
        \node[inner sep=0pt] at (3,2.5) {\includegraphics[width=.3\textwidth]{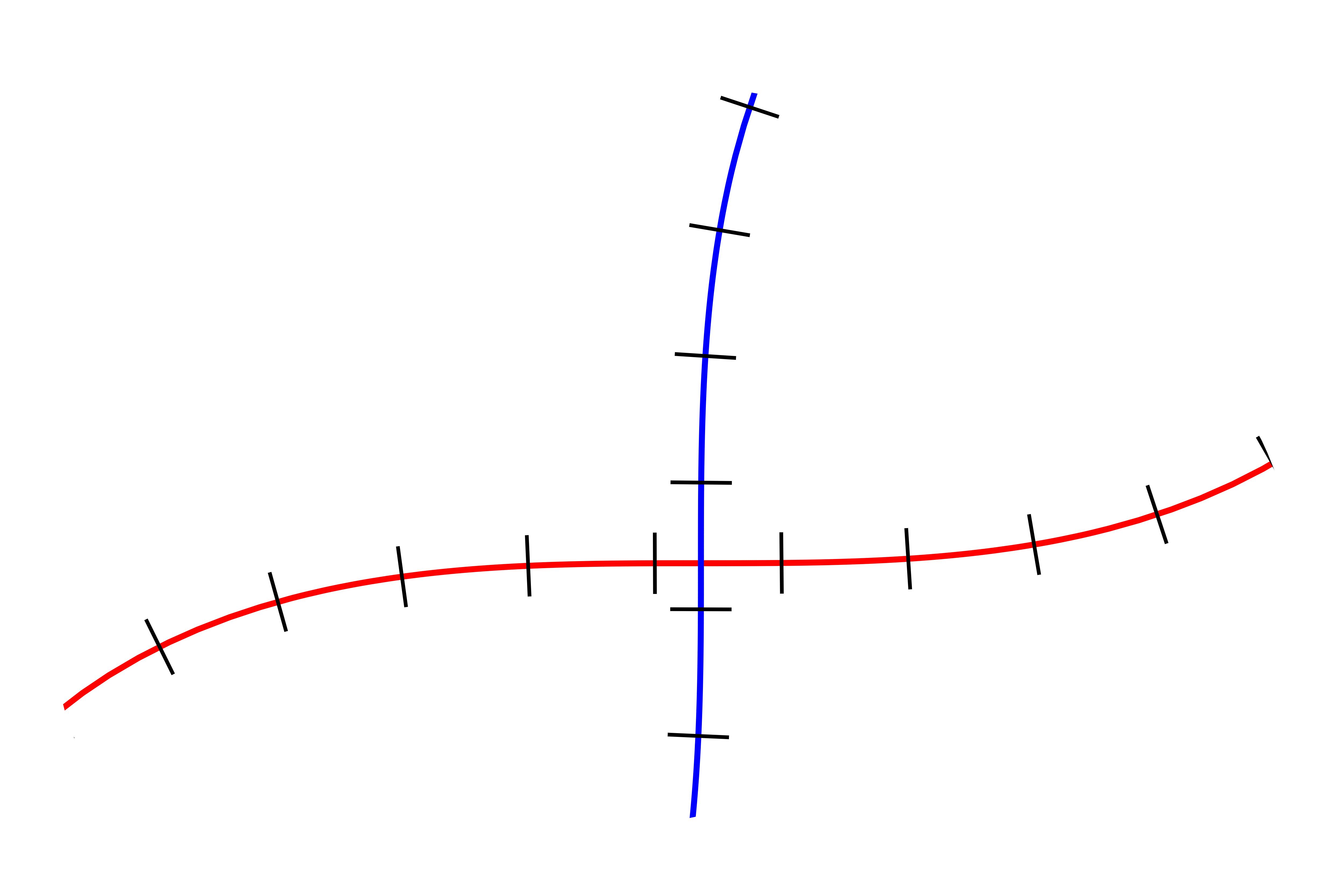}};
        \node[inner sep=0pt] at (5.35,1.9) {\includegraphics[width=.2\textwidth]{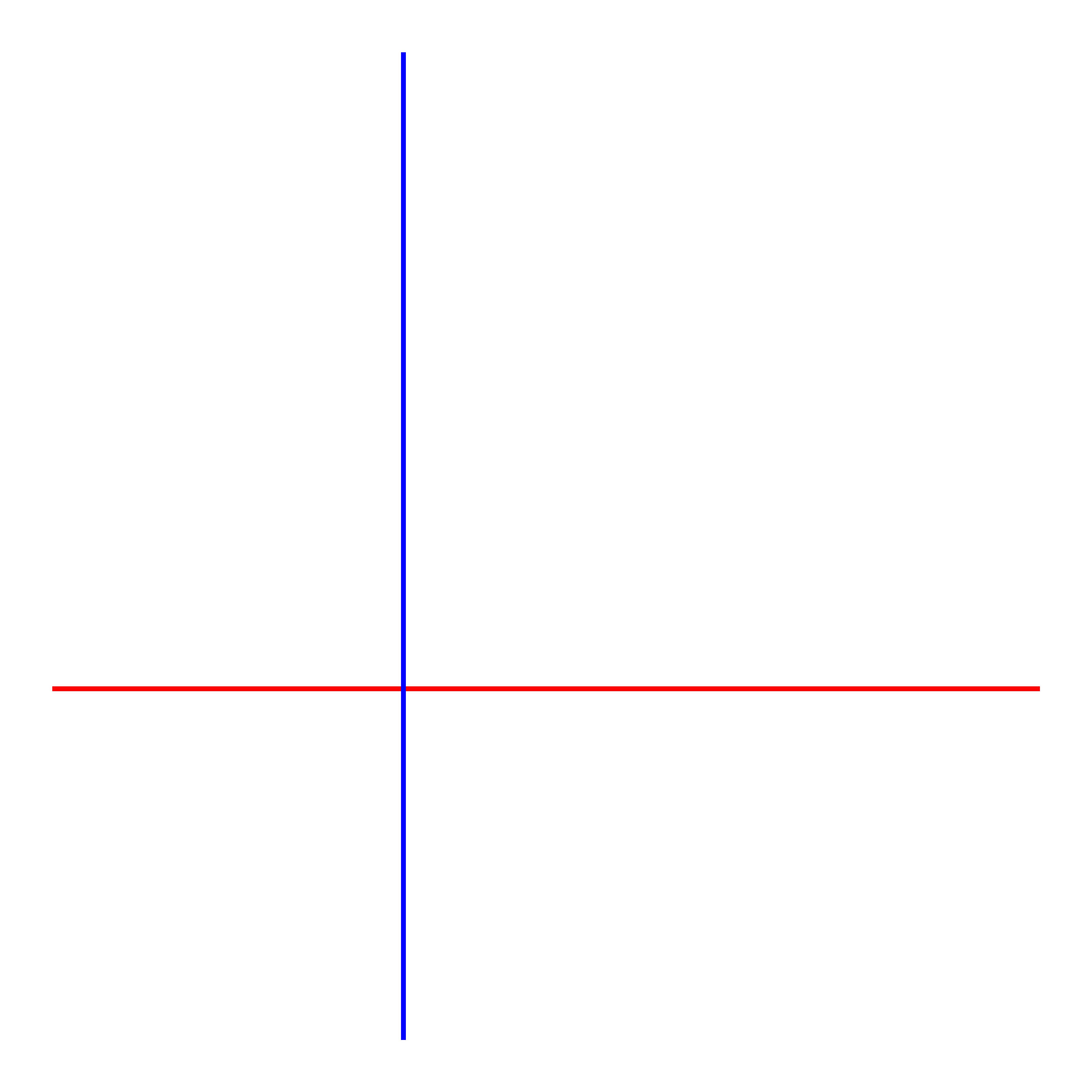}};
        \node[inner sep=0pt] at (4.02,0) {\includegraphics[width=.2\textwidth]{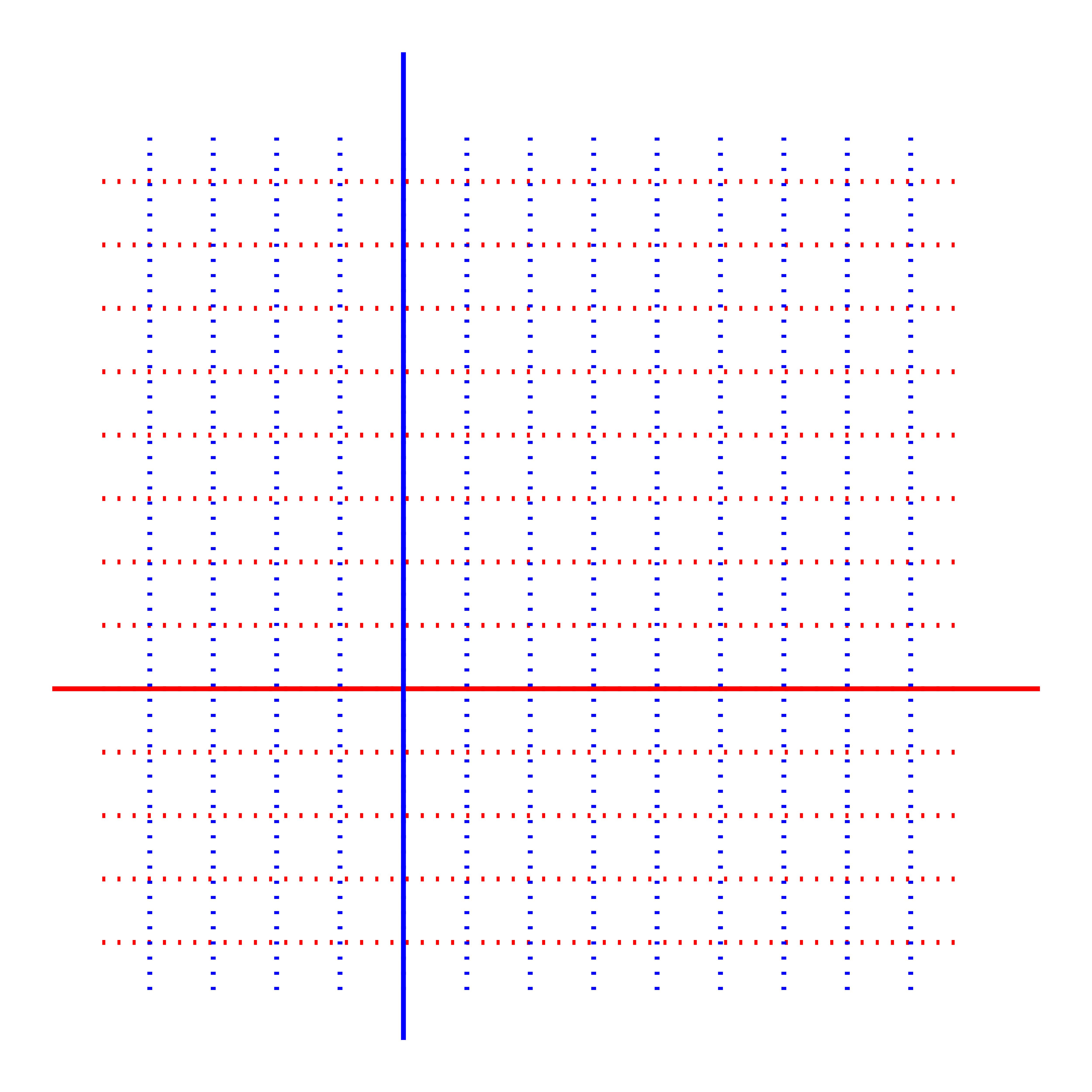}};
        \node[inner sep=0pt] at (0.85,0.35)   {\includegraphics[width=.3\textwidth]{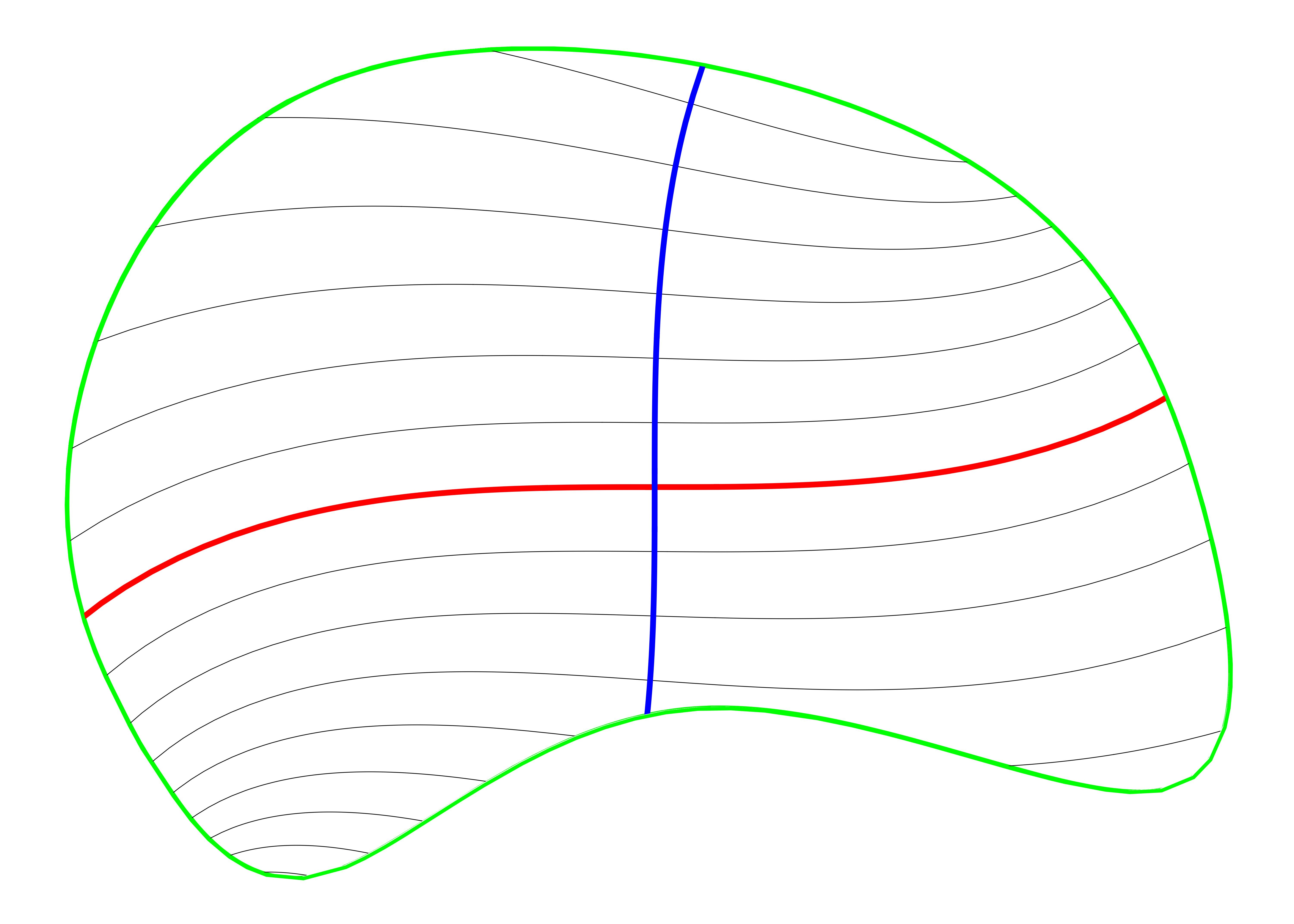}};
        \draw[->] (-0.925,1.75) to  (-0.925+0.4,1.75);   
        \draw[->] (-0.925,1.75) to  (-0.925,1.75+0.4);   
        \node[] at (-0.925+0.5,1.75) {$x$}; 
        \node[] at (-0.925,1.75+0.5) {$y$}; 
        \draw[->] (1.5+0.4,1.75) to  (1.5+0.4+0.4,1.75);   
        \draw[->] (1.5+0.4,1.75) to  (1.5+0.4,1.75+0.4);   
        \node[] at (1.5+0.5+0.4,1.75) {$x$}; 
        \node[] at (1.5+0.4,1.75+0.5) {$y$}; 
        \draw[->] (4.55,1.15) to  (4.55+0.4,1.15);   
        \draw[->] (4.55,1.15) to  (4.55,1.15+0.4);   
        \node[] at (4.55+0.5,1.11) {$u$}; 
        \node[] at (4.55,1.15+0.5) {$v$}; 
        \draw[->] (3.3,-0.65) to  (3.3+0.4,-0.65);   
        \draw[->] (3.3,-0.65) to  (3.3,-0.65+0.4);   
        \node[] at (3.3+0.5,-0.69) {$u$}; 
        \node[] at (3.3,-0.65+0.5) {$v$}; 
        \draw[->] (-0.3,-0.5) to  (-0.3+0.4,-0.5);   
        \draw[->] (-0.3,-0.5) to  (-0.3,-0.5+0.4);   
        \node[] at (-0.3+0.5,-0.5) {$x$}; 
        \node[] at (-0.3,-0.5+0.5) {$y$}; 
        \draw [-{Stealth[scale=2]},thick] (1.25,3) [out=10,in=175] to (2,3);
        \node[] at (1.5,3.12) {$\text{Gauge}$}; 
        \draw [-{Stealth[scale=2]},thick] (4,3) [out=5,in=120] to (4.8,2.6);
        \node[] at (4.5,3.1) {$\text{Goursat problem}$};
        \draw [-{Stealth[scale=2]},thick] (5.6,1) [out=-100,in=10] to (4.7,0.4);
        \node[] at (5.1,0.7) {$\text{Solution}$};
        \draw [-{Stealth[scale=2]},thick] (3.15,0.38) [out=187,in=-15] to (2.15,0.45);   
        \node[] at (2.85,0.2) {$\text{Map to }(x,y)$};
        \node[] at (-0.2,3.2) {$\text{Input}$}; 
        \node[] at (0.75,1.25) {$\text{Solution}$}; 
        \node[] at (3.5,2.52) {$^{\alpha_0(u)}$}; 
        \draw [|-|] (3.37,2.45) to (3.57,2.45);
        \node[] at (3.6,2.2) {$^{\kappa_g{(u)}}$}; 
        \node[] at (3.3,2.9) {$^{\kappa_g{(v)}}$}; 
        \node[] at (2.8,2.97) {$^{\beta_0(v)}$}; 
        \draw [|-|] (2.98,2.9) to (3.025,3.08);
        \node[] at (5.7,1.6) {$^{\alpha_0(u),\:\: r(u)}$}; 
        \node[] at (4.95,2.2) {$^{\beta_0(v)}$};
        \node[] at (5,2.07) {$^{t(v)}$}; 
        \node[] at (4.2,0.4) {$\alpha(u,v)$};
        \node[] at (4.2,0.25) {$\beta(u,v)$};
        \node[] at (0.85,1.8) {(I)};
        \node[] at (3.7,1.8) {(II)};
        \node[] at (5.9,1.2) {(III)};
        \node[] at (4.5,-0.85) {(IV)};
        \node[] at (1.7,-0.5) {(V)};
    \end{tikzpicture}
    \caption{Solving the PLCE director field. (I) A domain and two curves that intersect orthogonally are set in the laboratory $xy$ coordinates; the curves will become director and director and director perpendicular integral curves. (II) We fix the gauge functions $\alpha(u,v_0),\:\beta(u_0,v)$ by choosing a parameterization of the initial curves. The geodesic curvatures further fix the gauge functions $r(u),\:t(v)$. (III) This sets the Goursat initial value problem, whose (IV) solution in the $uv$ coordinates is given by the expression (\ref{eq:sol-alpha}). (V) Finally, we map the resolved director field back to the $xy$ coordinates, and restrict it to the desired domain.}
    \label{algorithm} 
\end{figure*}  

\paragraph*{Solution-}
The system in Eqs.~(\ref{eq:I:absy}) is in general genuinely nonlinear, however when $\mathcal{K}=0$ it reduces to
\begin{align}
    \frac{\partial}{\partial v}\left(\frac{1}{\beta} \frac{\partial \alpha }{\partial v}\right) & =\frac{\partial }{\partial u}\left(\frac{1}{\alpha}\frac{\partial \beta}{\partial u} \right)=0,
\end{align}
and could readily be integrated once to read
\begin{align}
    \frac{\partial \alpha(u,v)}{\partial v} =r(u)\beta(u,v),\quad
    \frac{\partial \beta(u,v)}{\partial u} =t(v)\alpha(u,v),
    \label{eq:II:ab}
\end{align}
with $r(u)$ and $t(v)$ arbitrary functions. Comparing with Eq.~(\ref{eq:I:bs-def}) reveals that these functions are not independent of our previous gauge choice, since
\begin{align}
b(u,v)=-\frac{r(u)}{\alpha(u,v)},\quad
s(u,v)=\frac{t(v)}{\beta(u,v)}.
\label{eq:I:rel}
\end{align}
The bend $b\left(u,v_0\right)$ and splay $s\left(u_0,v\right)$ are simply the geodesic curvatures of the director and director-perpendicular initial curves, respectively, setting an algebraic relation between $r(u),t(v)$ and $\alpha_0\left(u\right),\beta_0\left(v\right)$. Importantly, relations (\ref{eq:I:rel}) hold not only at the initial curves, but everywhere within the solution domain.

For any choice of $r(u)$ and $t(v)$ we can find the solution to this linear Goursat problem using Riemann's method (full derivation in Supplemental Materials). In short, we find a convolution kernel (also known as a Riemann's function) based on the integrals
\begin{align}
   R(u)\equiv\int_{u_0}^{u}\mathrm{d}u'\:r(u'),\quad T(v)\equiv\int_{v_0}^{v}\mathrm{d}v'\:t(v'). 
   \label{eq:II:RT}
\end{align}
The solution is then given by
\begin{widetext}
\begin{subequations}
\begin{align}
    \frac{\alpha(u,v)-\alpha_0(u)}{r(u)}&= \int_{u_0}^u \mathrm{d} u'\:\alpha_0(u')\sqrt{\frac{T(v)}{R(u)-R(u')}}I_1\left(2\sqrt{\left[R(u)-R(u')\right]T(v)}\right)+\int_{v_0}^v \mathrm{d}  v'\: \beta_0(v')I_0\left(2\sqrt{R(u)\left[T(v)-T(v')\right]}\right),\\    \frac{\beta(u,v)-\beta_0(v)}{t(v)}&= \int_{u_0}^u \mathrm{d} u'\:\alpha_0(u')I_0\left(2\sqrt{\left[R(u)-R(u')\right]T(v)}\right)+\int_{v_0}^v \mathrm{d}  v'\: \beta_0(v')\sqrt{\frac{R(u)}{T(v)-T(v')}}I_1\left(2\sqrt{R(u)\left[T(v)-T(v')\right]}\right),
\end{align}\label{eq:sol-alpha}
\end{subequations}
\end{widetext}
with $I_n$ the modified Bessel function of order $n$.

Of course, we are interested not in the scale factor functions $\alpha$ and $\beta$, but rather in the director field $\theta(x,y)$. Eq.~(\ref{eq:I:rel}) implies that $r(u)=-\partial_u\theta(u,v)$ and $t(v)=\partial_v\theta(u,v)$, thus the change in $\theta$ along $u-$lines is independent of the value of $v$ and vice versa.
Combined with Eq.~(\ref{eq:II:RT}), we obtain
\begin{align}
  \theta(u,v)-\theta_0=T(v)-R(u),
\label{eq:II:dir}
\end{align}
with $\theta_0$ an arbitrary constant. To obtain the solution in the laboratory Cartesian coordinates we plug the solutions in Eqs.~(\ref{eq:sol-alpha},\ref{eq:II:dir}) to the $xy-uv$ transformation defined by Eq.~(\ref{eq:I:eq:embe}), thus
\begin{align}\label{eq:II:lab-sys}
\mathbf{r}(u,v)-\mathbf{r}_0&=\int_{u_0}^u\mathrm{d}u'\:\alpha(u',v_0)\:\hat{\mathbf{n}}(u',v_0)\nonumber\\&+\int_{v_0}^v\mathrm{d}v'\:\beta(u,v')\:\hat{\mathbf{n}}_\perp\left(u,v'\right).
\end{align}
Together, Eqs.~(\ref{eq:II:dir}, \ref{eq:II:lab-sys}) provide us with $x\left(u,v\right)$, $y\left(u,v\right)$ and $\theta(u,v)$, from which one extracts $\theta(x,y)$ and can go on to make their PLCE. The algorithm is illustrated in the Fig.~\ref{algorithm}. An initial domain and two curves that intersect each other orthogonally are chosen. In the $uv-$plane, these curves become straight lines, and the solution away from those lines is given by Eq.~(\ref{eq:sol-alpha}). With Eqs.~(\ref{eq:II:dir}, \ref{eq:II:lab-sys}), we map back the solution to the input domain in lab coordinates.

\paragraph*{Singularities-}
A solution cannot be further extended beyond a point where either $\alpha=0$ or $\beta=0$, and the PDE system becomes singular. At these points gradients of the nematic director diverge, namely these points are disclinations. Gevirtz \cite{gevirtz01} proved that, even though there is no bound for the number of singularities that can appear in a given domain, they are of only two types. In the language of nematic liquid crystals, these types correspond to very specific realizations of a $+1$ and a $+1/2$ topological defects. The $+1$ type has a logarithmic spiral shaped director \cite{mostajeran2,modes-warner}. The $+1/2$ type is made of a purely azimuthal sector and a purely radial sector, separated by two $\pi/2$ constant-director sectors. In both cases, these structures would generically upon actuation make a cone or an anti-cone, the opening angle of which depends on the spiral/sector angle. However, if one sets the spiral/sector angle just right, such singular LCEs will make neither a cone nor an anti-cone. They will deform in the plane but remain flat everywhere, including at the defect apex.

This highly non-trivial result extends beyond PLCEs (since for any bounded Gaussian curvature, at small enough distances $r\ll|K_\mathsf{A}|^{-1/2}$ the surface appears nearly flat). Therefore, a point disclination in an LCE sheet would generically induce a diverging Gaussian curvature near or at the defect apex upon actuation, unless it is locally one of the two above-mentioned director fields near the tip. Thus, a \emph{smooth} LCE sheet that is to remain \emph{smooth} upon actuation may only include $+1$ and a $+1/2$ topological defects. 

\paragraph*{Gauge choice-} 
A natural gauge choice, that grossly simplifies the integral solution (\ref{eq:sol-alpha}), is to set $r(u)=t(v)=-1$. This gauge, which we call the Hencky-Prandtl ($\textsf{HP}$) gauge for reasons that will become apparent below, is widely used in different contexts in the mathematical literature \cite{Collins,johnson-book,Graczykowski}. In the HP gauge, Eq.~(\ref{eq:II:ab}) becomes the Klein-Gordon equation for both $\alpha$ and $\beta$:
\begin{align}
    \frac{\partial^2\alpha^{\mathsf{HP}}}{\partial u\partial v}=\alpha^{\mathsf{HP}},
    \label{eq:II:kg}
\end{align}
In this gauge we have that
\begin{align}
    \alpha^{\mathsf{HP}}(u,v)=\frac{1}{b(u,v)},\quad\beta^{\mathsf{HP}}(u,v)=-\frac{1}{s(u,v)},
\end{align}
so that $\alpha,\:\beta$ correspond to the radii of curvature of the $u-$ and $v-$lines at any point.

In many cases, if $\alpha_0(u)$ and $\beta_0(v)$ are simple enough, Eq.~(\ref{eq:sol-alpha}) could be integrated explicitly. In particular, Taylor-expanding $\alpha_0(u)$ and $\beta_0(v)$ with coefficients $\alpha_n$ and $\beta_n$ respectively, we obtain a power-series solution (see Supplemental Materials):
\begin{align}
    &\alpha(u,v)=\alpha(u_0,v_0)I_0\left(2\sqrt{(u-u_0)(v-v_0)} \right)\label{eq:series-sol}\\
    &+\sum_{n=1}^\infty \left[\alpha_n\left(\frac{u-u_0}{v-v_0} \right)^{\frac{n}{2}}-\beta_{n-1}\left(\frac{v-v_0}{u-u_0} \right)^{\frac{n}{2}} \right] \nonumber\\
    &\:\:\:\:\:\:\:\:\:\:\:\:\:\:\times I_{n}\left(2\sqrt{(u-u_0)(v-v_0)} \right).\nonumber
\end{align} 

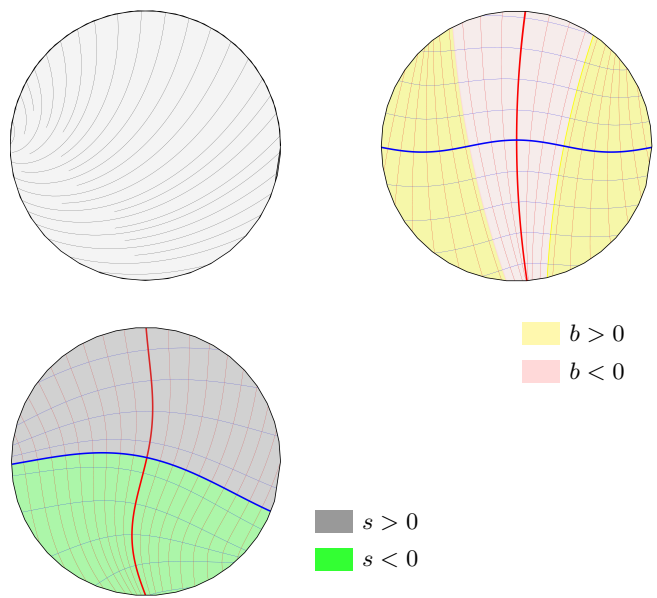
\begin{figure}
    \centering 
    \begin{tikzpicture}
        \node[inner sep=0pt] at (-4,1) {\includegraphics[width=.2\textwidth]{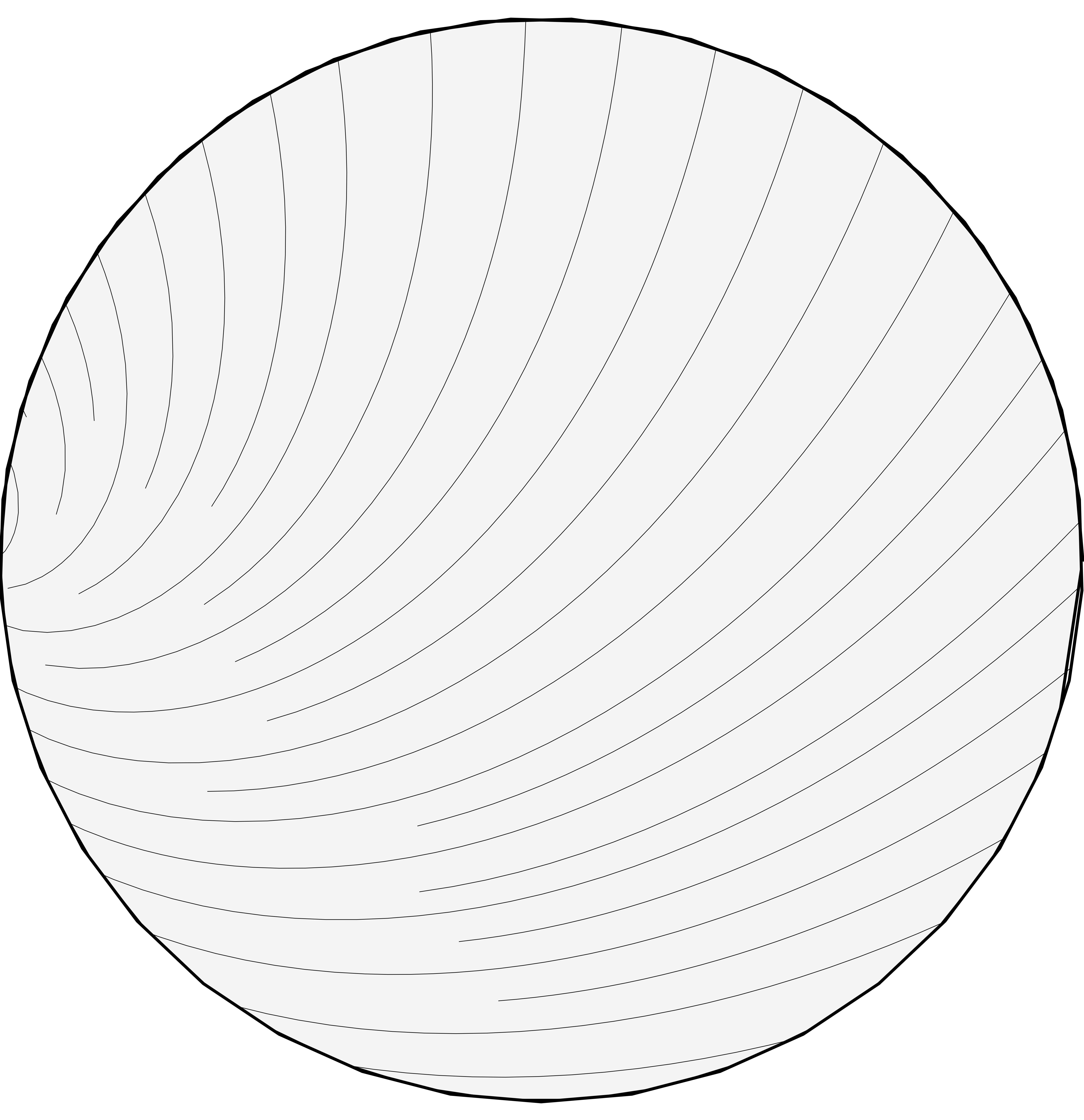}}; 
        \node[inner sep=0pt] at (1,1) {\includegraphics[width=.21\textwidth]{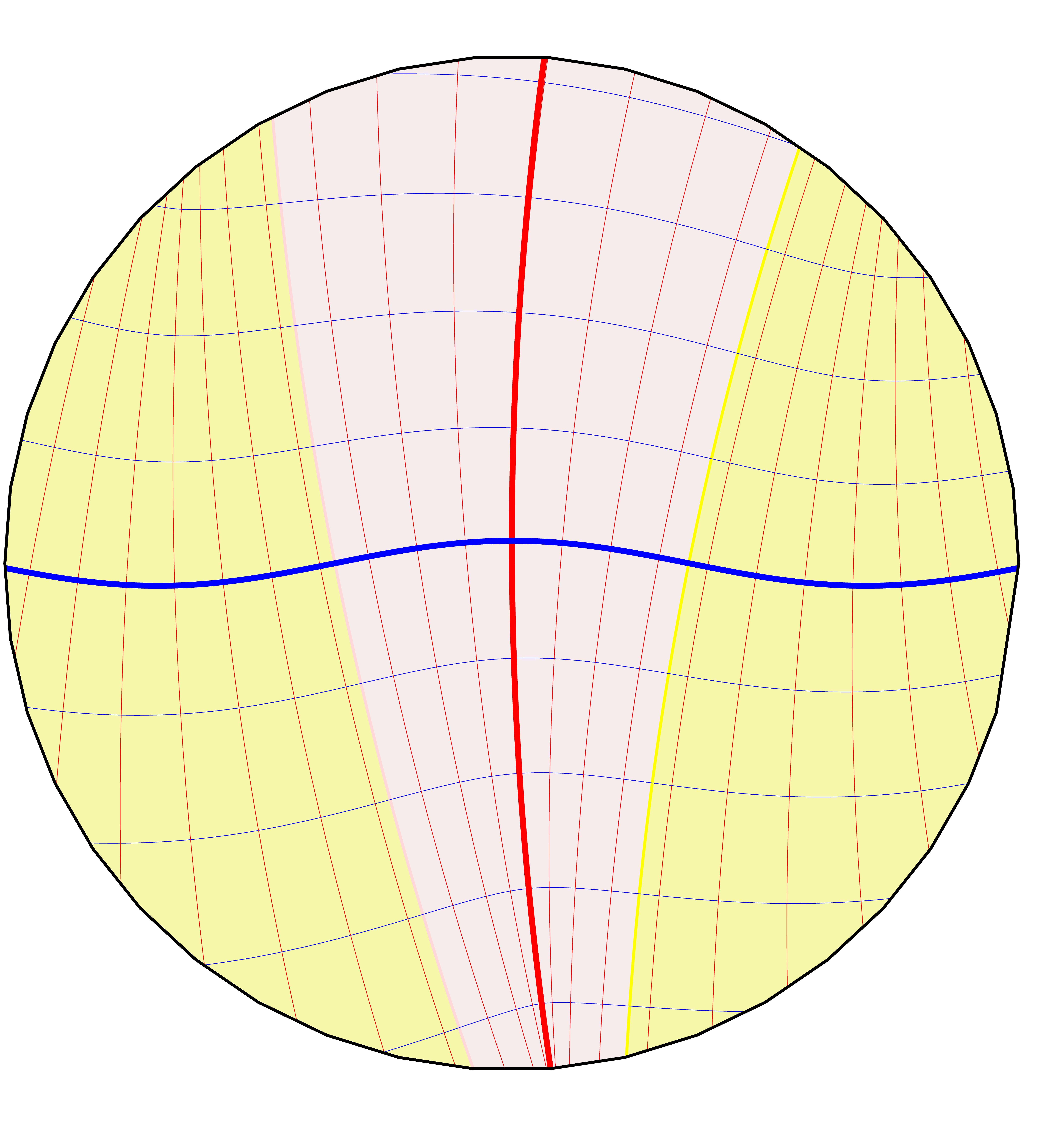}};
        \node[inner sep=0pt] at (-4,-3.2) {\includegraphics[width=.21\textwidth]{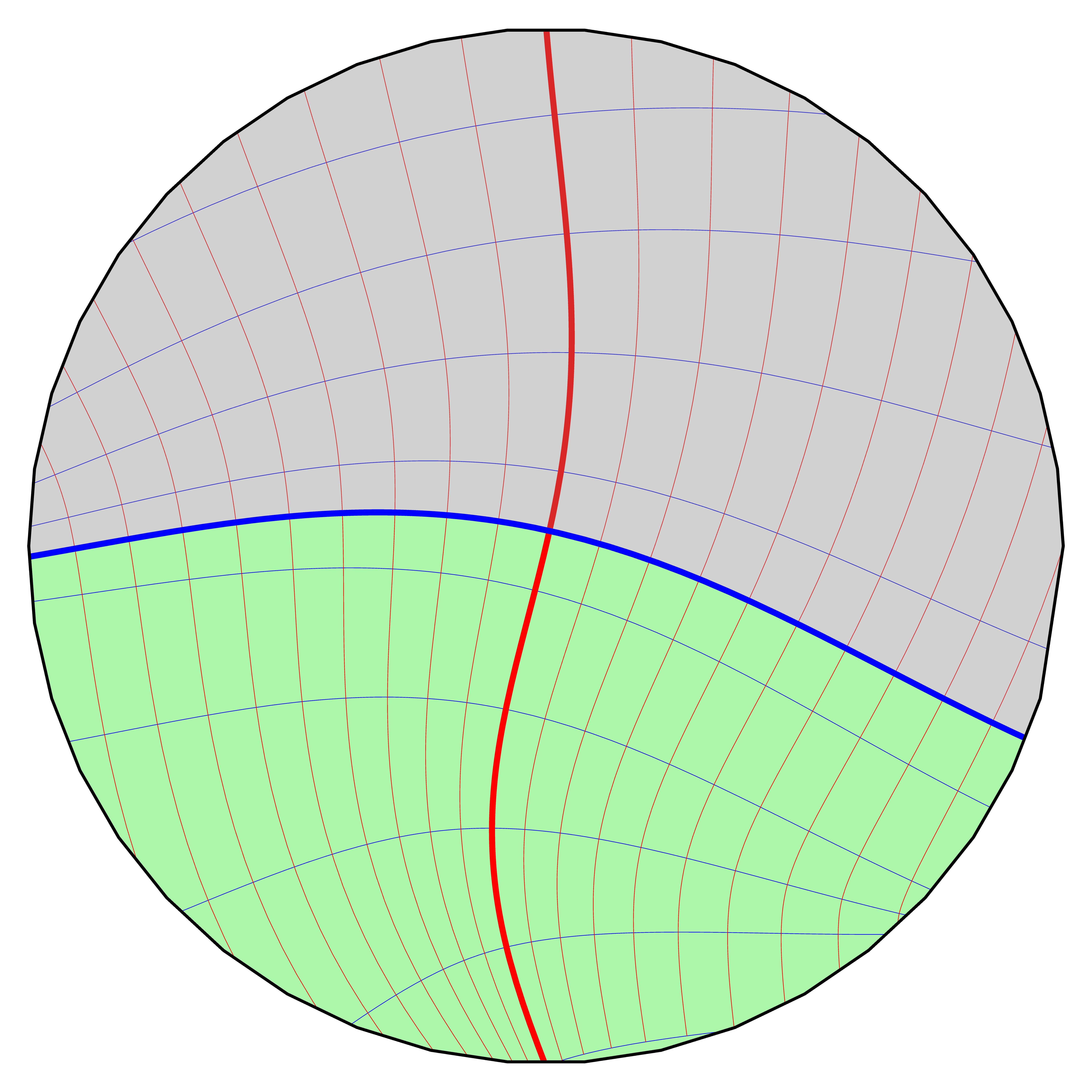}};
        \fill [yellow!40] (1.0,-1.65) rectangle (1.5,-1.35); 
        \fill [red!15] (1.0,-1.85) rectangle (1.5,-2.15);
        \node[] at (2,-1.5) {$b>0$}; 
        \node[] at (2,-2.0) {$b<0$};   
        \fill [gray!80] (-1.75,-4.15) rectangle (-1.25,-3.85); 
        \fill [green!80] (-1.75,-4.35) rectangle (-1.25,-3.85-0.8);
        \node[] at (-0.75,-4) {$s>0$};   
        \node[] at (-0.75,-4.5) {$s<0$};  
    \end{tikzpicture}
    \caption{Geodesic curvature changes of the director field integral curves. (Top-left) The nematic director is everywhere positively bent and positively splayed. The geodesic curvatures of both sets of integral curves are everywhere nonzero. (Top-right) Here, the bend $b$ changes sign twice while $s$ remains positive; the change of signs occurs along $v$ curves, a hallmark of LPCEs. (Bottom) Similarly, $s$ could only change sign across $u$ curves.}
    \label{example-signs} 
\end{figure}

Eq.~(\ref{eq:I:rel}) implies that, in a PLCE, if the nematic bend $b$ changes sign it does so across $v-$lines, and likewise, the splay $s$ only changes sign across $u-$lines. Regularity of the PDE system requires that in the HP gauge $b$ and $s$ do not change sign. Therefore, this gauge is only useful in cases where the initial curves' geodesic curvatures do not change sign and, as a result, the bend and splay are everywhere nonzero. Such an example is shown in Fig.~\ref{example-signs}(a). In cases where the curvature of the initial curves change sign, the HP gauge is rendered impracticable.

\begin{figure}
     \centering  
     \begin{tikzpicture}
        \node[inner sep=0pt] at (-4.05,0) {\includegraphics[width=.2\textwidth]{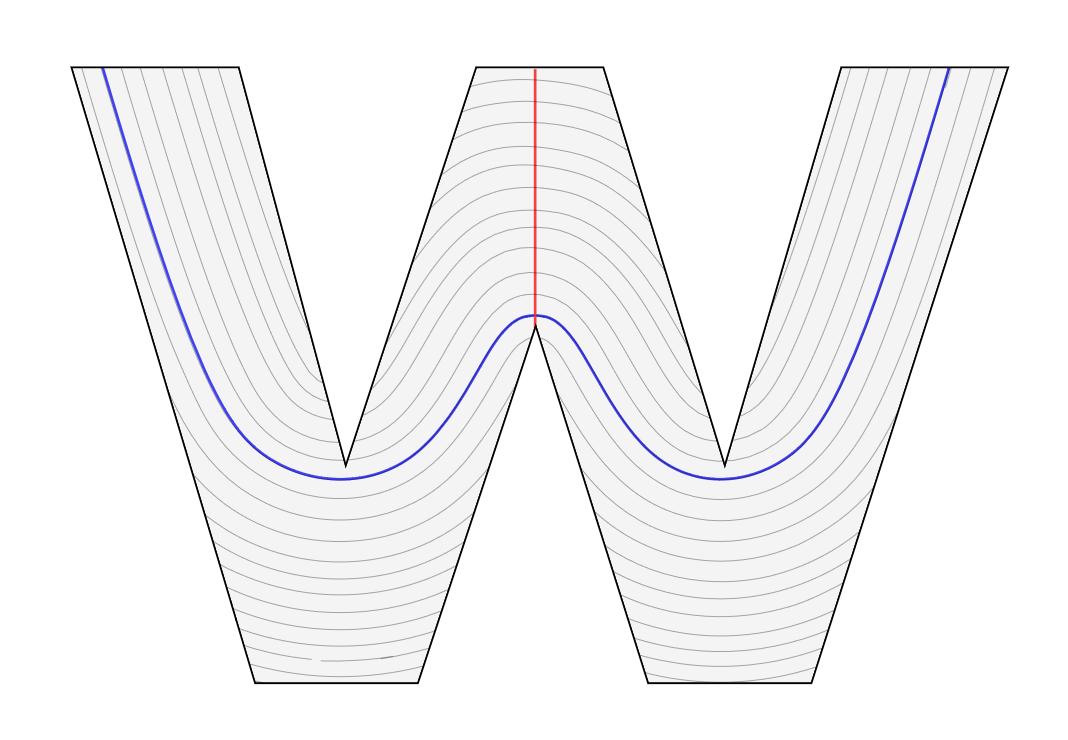}};  
        \node[inner sep=0pt] at (-2.02,0) {\includegraphics[width=.045\textwidth]{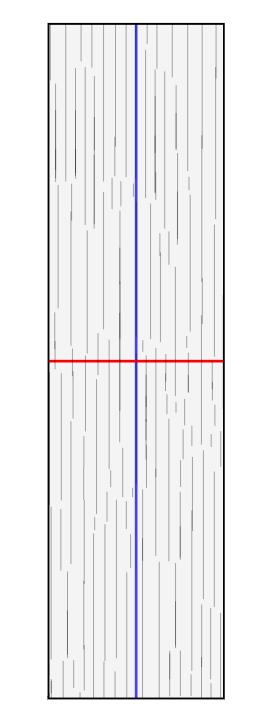}};   
        \node[inner sep=0pt] at (-0.8,0) {\includegraphics[width=.1\textwidth]{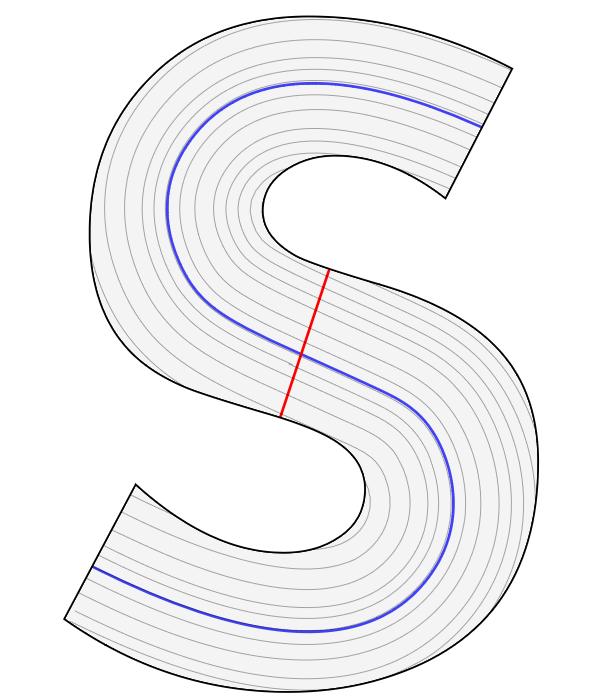}};  
        \node[inner sep=0pt] at (0.05,2.5) {\includegraphics[width=.35\textwidth]{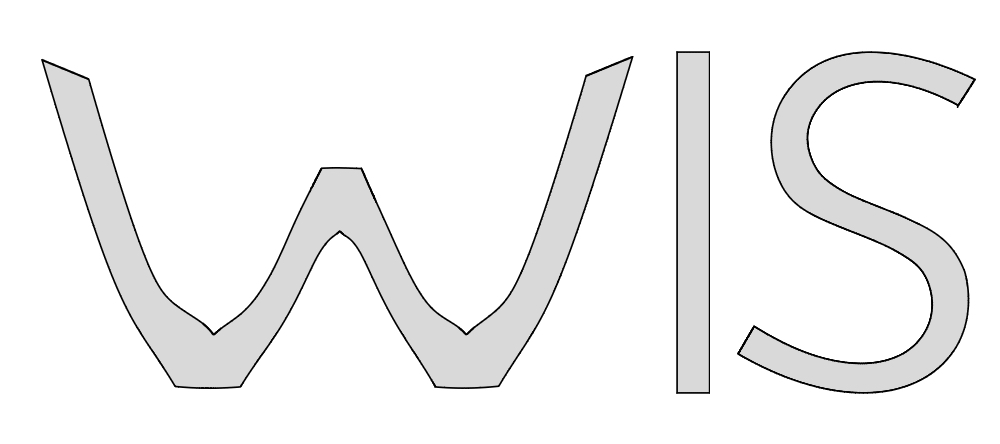}}; 
        \node[inner sep=0pt] at (0.5,-2.3) {\includegraphics[width=.27\textwidth]{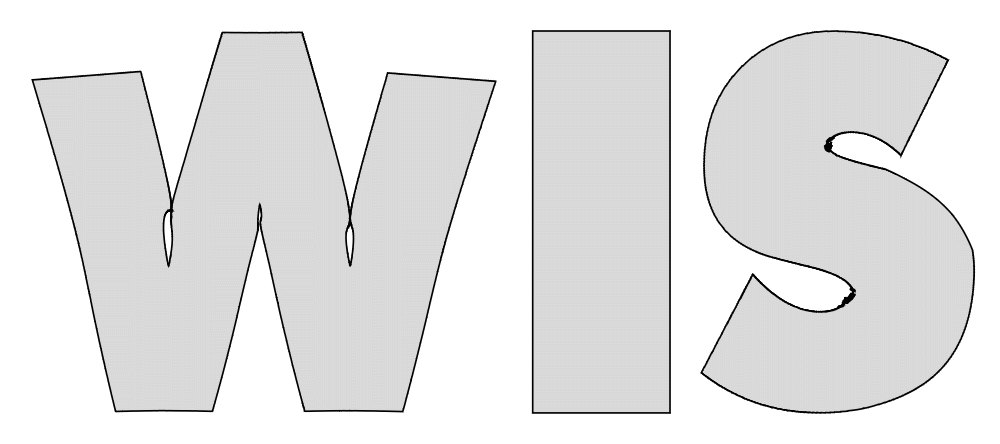}};
        \node[] at (-4.1,2.8) {$\lambda_1/\lambda_2=4$}; 
        \node[] at (-3.5,-2.8) {$\lambda_1/\lambda_2=0.6$}; 
        \draw [-{Stealth[scale=2]},thick] (-4.8,1.2) [out=80,in=180] to (-2.8,2.6);   
        \draw [-{Stealth[scale=2]},thick] (-3.75,-1.3) [out=-80,in=180] to (-2.0,-2.5);
    \end{tikzpicture}
    \caption{A PLCE designed to change the font weight of a text. The initial $u$ curves are chosen to run along the letters' backbones, while the $v$ curves are chosen to be straight lines. Actuation makes the font either lighter ($\lambda_2<\lambda_1$) or bolder ($\lambda_1<\lambda_2$), without buckling out of the plane.}
    \label{fig:wis}
\end{figure}
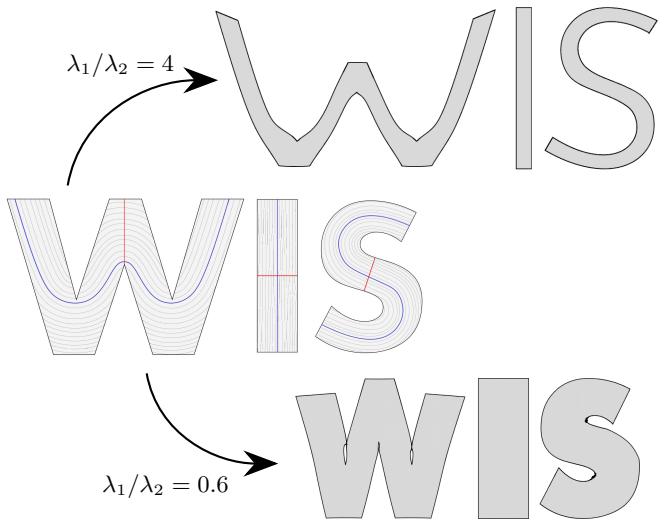

One way out is to generalize it so that $r(u)$ and $t(v)$ are simple polynomials. A more natural choice, suggested by Niv and Efrati \cite{niv-efrati} and used in \cite{griniasty-aharoni-efrati}, is to set $\alpha_0(u)=\beta_0(v)=1$ which is equivalent to
\begin{align}
    r^{\mathsf{NE}}(u)=-b(u,v_0),\quad
    t^{\mathsf{NE}}(v)=s(u_0,v),
\end{align}
and therefore, by Eq.~(\ref{eq:I:rel})
\begin{align}
    \alpha^{\mathsf{NE}}(u,v)=\frac{b(u,v_0)}{b(u,v)},\quad 
    \beta^{\mathsf{NE}}(u,v)=\frac{s(u_0,v)}{s(u,v)}.
\end{align}
With the Niv-Efrati gauge, explicit integration becomes harder and will often need to be carried out numerically. Nonetheless it allows designing PLCEs with bend and splay that change sign (Fig.~\ref{example-signs}(b-c), Fig.~\ref{fig:wis}).

\paragraph*{Discussion-}
It is worth mentioning that, due to the purely geometric nature of this problem, the system that we consider has analogues in other physical and engineering contexts \cite{acharya}. The networks of director and director-perpendicular fields that we obtain for PLCEs are known as Hencky-Prandtl (HP) nets \cite{Hencky, Prandtl}. They appear in the field of structural optimisation as load-carrying structures of minimum weight \cite{Michell, whittle, michell-book}, and in plasticity theory \cite{Collins, hill-book,johnson-book, Collins-plate} as slip line fields, that generate deformation patterns in plastic solids. In both cases, these solutions emerge due to the assumed-constant yield stress/strain of the material, analogues to the constant elongation factors $\lambda_{1,2}$ in the PLCE setting. Several numerical techniques for the construction of HP nets were developed in the applied mathematics and engineering literature; the analytical solution (\ref{eq:sol-alpha}) for the particular case of $r=t=-1$ was derived using classical methods by Geiringer \cite{Geiringer}, and employed in diverse geometries by various authors \cite{Arcisz, Thomason, Graczykowski}.

In the PLCE context, the Goursat problem we solve in this manuscript may at first glance seem artificial. However, as we show next, it is useful and may be applied in various ways for design and engineering purposes. Generic LCE sheets, that develop Gaussian curvature upon actuation \cite{aharoni2018universal,mostajeran2,modes-warner}, are inherently elastically frustrated; buckling out of the plane to realize their non-Euclidean geometry and reduce stretching will result in an elastic bending cost. At any nonzero thickness, upon actuation there exists no stress-free state for such sheet, even when one introduces director gradients across the thickness to reduce bending \cite{aharoni2018universal}. PLCEs, on the other hand, are inherently compatible and admit a stress-free state at any thickness and any actuation parameter level. Thus, the solutions presented here could be used for making bulk shape-shifting objects, e.g. long beams that are programmed to change their cross-section upon actuation.

Having a full analytical solution rather than an implicit system of equations opens the door to several analyses and applications that are not currently accessible. The singularity horizon could be more easily identified, and the problem of domain design more accessible (although, as mentioned in the introduction, the full inverse problem is not solvable). The analytical solution further provides tools for optimization problems over the set of solutions, handling singularities and more.

Finally, the Goursat initial conditions are particularly useful for design purposes, as one may arbitrarily choose the two initial curves to induce elongation/shortening along predefined paths. One example for a useful application of this design principle is shown in Fig.~\ref{fig:wis}. If one chooses the initial curve along the letter backbone of a certain typeface, the resulting PLCE will change its font weight upon actuation. As mentioned before, these font designs will be stress-free and morphologically robust regardless of their thickness. This and other simple design principles may build upon this work, and provide useful tools for future design and technology applications.

\begin{acknowledgments}
We thank Julian Gevirtz, Asaf Shachar and Raz Kupferman for useful discussions. This research was supported by the Israel Science Foundation (grant No. 2677/20).
\end{acknowledgments}

\bibliography{apssamp}

\newpage
\input{sm.tex}

\end{document}

%% file: sm.tex



	\appendix
	
	\section{Integration of the Riemann function} 
	
	The system $\left(\partial^2_{uv}-f(u,v) \right)\mathcal{R}=0$, subject to $\mathcal{R}\left(u_0,v;u_0,v_0 \right)=\mathcal{R}\left(u,v_0;u_0,v_0 \right)=1$ is equivalent to the integral equation 
	
	\begin{align}
		\mathcal{R}(u,v;u_0,v_0)=1+\int_{u_0}^u\mathrm{d} u'\int_{v_0}^v\mathrm{d} v'\:f\left(u',v'\right)\mathcal{R}\left(u',v';u_0,v_0 \right),
	\end{align}
	
	and successively adjoining it into itself we have a series of nested integrals:
	
	\begin{align} 
		& \mathcal{R}(u,v;u_0,v_0)\\
		&=1+\int_{uv} \mathrm{d} u'\mathrm{d} v'\int_{u'v'}\mathrm{d} u''\mathrm{d} v''\cdots\int_{u^{(k-1)}v^{(k-1)}}\mathrm{d} u^{(k)}\mathrm{d} v^{(k)}\nonumber\\
		&\:\:\:\:\:\:\:\:\:\:\:\:\:\:\times f\left(u',v'\right)f\left(u'',v''\right)\cdots f\left(u^{(k)},v^{(k)}\right) +  \cdots.\nonumber 
	\end{align} 
	
	If $f\left(u,v\right)=t\left(v\right)r\left(u\right)$, then the $u$ and $v$ dependence is factorized on each term of the sum, that is, the $k^\text{th}$-term is

	\begin{align}
		&\int_{u_0}^u \mathrm{d} u'\cdots\int^{u^{(k-1)}}_{u_0}\mathrm{d} u^{(k)}\: r\left(u'\right)\cdots r\left(u^{(k)}\right)\\
		&\times \int_{v_0}^v \mathrm{d} v'\cdots\int^{v^{(k-1)}}_{u_0}\mathrm{d} v^{(k)}\: t\left(u'\right)\cdots t\left(v^{(k)}\right)\nonumber\\
		&=\frac{1}{\left(k!\right)^2}  \left(  \int_{u_0}^u \mathrm{d} u'\:r\left(u'\right) \int_{v_0}^v \mathrm{d} v'\:t\left(v'\right)\right)^k.
	\end{align}
	
	The factorial squared power series equals the modified Bessel function with $2z^{1/2}$ argument, so we can compactly write 
	
	\begin{align}
		\mathcal{R}\left(u,v;u_0,v_0\right)=I_0\left(2\left(R\left(u_0,u\right)T\left(v_0,v\right)\right)^{1/2}\right).\nonumber
	\end{align}
	
	\section{Derivation of the series solution}
	
	Using, for simplicity, the $\mathsf{HP}$ gauge where $ \mathcal{R}(x,y;u,v)=I_0\left(2\left((x-u)(y-v)\right)^{1/2}\right)$, and setting $u_0=v_0=0$, without loss of generality, the solution (9) for $\alpha(u,v)$ can be written as
	
	\begin{align}
		\alpha(u,v)=&\alpha(u_0,v_0)I_0\left(2\sqrt{uv}\right)+\int_{0}^u\:\mathrm{d} t\frac{\mathrm{d} \alpha(t,0) }{\mathrm{d} t}I_0\left(2\sqrt{v(u-t)}\right)\nonumber\\
		&-\int_{0}^v\mathrm{d} t\: \beta(0,t) I_0\left(2\sqrt{u(v-t)}\right).
		\label{eq:II:ser1}
	\end{align}
	
	We consider an interval where the Taylor series for $\alpha(x,0)$ and  $\beta(0,x)$ are convergent, and setting the origin at some common point, we can write
	
	\begin{align}
		\alpha(x,0)=\sum_{n=0}^\infty \frac{\alpha_n}{n!}\:x^n\:\text{ and }\: \beta(0,x)=\sum_{n=0}^\infty \frac{\beta_n}{n!}\:x^n,\nonumber
	\end{align}
	
	thus using the following convenient representation for the Riemann function 
	
	\begin{align}
		I_0\left(2\sqrt{uv}\right)=\int_{-\infty}^{(0^+)}\frac{\mathrm{d}\rho}{2\pi i \rho} \: e^{u\rho+v/\rho},
	\end{align}
	
	(the contour runs from $-\infty$ below the real axis and encircles the origin in the counterclockwise direction, going back to $-\infty$ above the axis) and replacing back in the integrals in (\ref{eq:II:ser1}) we have for the first one 
	
	\begin{align}
		& \int_{0}^u\:\mathrm{d} t\frac{\mathrm{d} \alpha(t,0) }{\mathrm{d} t}I_0\left(2\sqrt{v(u-x)}\right)    \label{eq:II:inte}
		\\ 
		&=\sum_{n=0}^\infty \frac{\alpha_{n+1}}{n!}\int_{-\infty}^{(0^+)}\frac{\mathrm{d} \rho}{2\pi i \rho} \: e^{v\rho+u/\rho}\int_0^u\mathrm{d} x\:x^n\:e^{-x/\rho}\nonumber.
	\end{align} 
	
	The last integral equals $\rho^{n+1}\left( \Gamma(n+1)-\Gamma(n+1,u/\rho)\right)$, with the second $\Gamma$ being the incomplete function, which can be expanded in positive powers so its contour integral vanishes and we have that
	
	\begin{subequations}
		\begin{align}
			\text{\ref{eq:II:inte}}&= \sum_{n=0}^\infty\alpha_{n+1}\int_{-\infty}^{(0^+)}\frac{\mathrm{d} \rho}{2\pi i \rho} \: e^{v\rho+u/\rho}\rho^n\\
			&=\sum_{n=0}^\infty\alpha_{n+1}\frac{\partial^{n+1}}{\partial v^{n+1}}I_0\left(2\sqrt{uv} \right)\\
			&=\sum_{n=0}^\infty\alpha_{n+1}\left(\frac{u}{v} \right)^{\frac{n+1}{2}}I_{n+1}\left(2\sqrt{uv} \right).
		\end{align}
	\end{subequations}
	
	For the second integral in (\ref{eq:II:ser1}) an identical calculation follows, and we can write down the expression (14) in the main text.

%% file: main.bbl
\providecommand{\noopsort}[1]{}\providecommand{\singleletter}[1]{#1}%
\begin{thebibliography}{36}%
\makeatletter
\providecommand \@ifxundefined [1]{%
 \@ifx{#1\undefined}
}%
\providecommand \@ifnum [1]{%
 \ifnum #1\expandafter \@firstoftwo
 \else \expandafter \@secondoftwo
 \fi
}%
\providecommand \@ifx [1]{%
 \ifx #1\expandafter \@firstoftwo
 \else \expandafter \@secondoftwo
 \fi
}%
\providecommand \natexlab [1]{#1}%
\providecommand \enquote  [1]{``#1''}%
\providecommand \bibnamefont  [1]{#1}%
\providecommand \bibfnamefont [1]{#1}%
\providecommand \citenamefont [1]{#1}%
\providecommand \href@noop [0]{\@secondoftwo}%
\providecommand \href [0]{\begingroup \@sanitize@url \@href}%
\providecommand \@href[1]{\@@startlink{#1}\@@href}%
\providecommand \@@href[1]{\endgroup#1\@@endlink}%
\providecommand \@sanitize@url [0]{\catcode `\\12\catcode `\$12\catcode
  `\&12\catcode `\#12\catcode `\^12\catcode `\_12\catcode `\%12\relax}%
\providecommand \@@startlink[1]{}%
\providecommand \@@endlink[0]{}%
\providecommand \url  [0]{\begingroup\@sanitize@url \@url }%
\providecommand \@url [1]{\endgroup\@href {#1}{\urlprefix }}%
\providecommand \urlprefix  [0]{URL }%
\providecommand \Eprint [0]{\href }%
\providecommand \doibase [0]{https://doi.org/}%
\providecommand \selectlanguage [0]{\@gobble}%
\providecommand \bibinfo  [0]{\@secondoftwo}%
\providecommand \bibfield  [0]{\@secondoftwo}%
\providecommand \translation [1]{[#1]}%
\providecommand \BibitemOpen [0]{}%
\providecommand \bibitemStop [0]{}%
\providecommand \bibitemNoStop [0]{.\EOS\space}%
\providecommand \EOS [0]{\spacefactor3000\relax}%
\providecommand \BibitemShut  [1]{\csname bibitem#1\endcsname}%
\let\auto@bib@innerbib\@empty
\bibitem [{\citenamefont {Dervaux}\ and\ \citenamefont
  {Ben~Amar}(2008)}]{dervaux-ben-amar}%
  \BibitemOpen
  \bibfield  {author} {\bibinfo {author} {\bibfnamefont {J.}~\bibnamefont
  {Dervaux}}\ and\ \bibinfo {author} {\bibfnamefont {M.}~\bibnamefont
  {Ben~Amar}},\ }\bibfield  {title} {\bibinfo {title} {Morphogenesis of growing
  soft tissues},\ }\href@noop {} {\bibfield  {journal} {\bibinfo  {journal}
  {Phys. Rev. Lett.}\ }\textbf {\bibinfo {volume} {101}},\ \bibinfo {pages}
  {068101} (\bibinfo {year} {2008})}\BibitemShut {NoStop}%
\bibitem [{\citenamefont {Armon}\ \emph {et~al.}(2011)\citenamefont {Armon},
  \citenamefont {Efrati}, \citenamefont {Kupferman},\ and\ \citenamefont
  {Sharon}}]{armon-efrati-kupferman}%
  \BibitemOpen
  \bibfield  {author} {\bibinfo {author} {\bibfnamefont {S.}~\bibnamefont
  {Armon}}, \bibinfo {author} {\bibfnamefont {E.}~\bibnamefont {Efrati}},
  \bibinfo {author} {\bibfnamefont {R.}~\bibnamefont {Kupferman}},\ and\
  \bibinfo {author} {\bibfnamefont {E.}~\bibnamefont {Sharon}},\ }\bibfield
  {title} {\bibinfo {title} {Geometry and mechanics in the opening of chiral
  seed pods},\ }\href {https://doi.org/10.1126/science.1203874} {\bibfield
  {journal} {\bibinfo  {journal} {Science}\ }\textbf {\bibinfo {volume}
  {333}},\ \bibinfo {pages} {1726} (\bibinfo {year} {2011})}\BibitemShut
  {NoStop}%
\bibitem [{\citenamefont {Klein}\ \emph {et~al.}(2007)\citenamefont {Klein},
  \citenamefont {Efrati},\ and\ \citenamefont {Sharon}}]{Klein-Efrati-Sharon}%
  \BibitemOpen
  \bibfield  {author} {\bibinfo {author} {\bibfnamefont {Y.}~\bibnamefont
  {Klein}}, \bibinfo {author} {\bibfnamefont {E.}~\bibnamefont {Efrati}},\ and\
  \bibinfo {author} {\bibfnamefont {E.}~\bibnamefont {Sharon}},\ }\bibfield
  {title} {\bibinfo {title} {Shaping of elastic sheets by prescription of
  non-euclidean metrics},\ }\href@noop {} {\bibfield  {journal} {\bibinfo
  {journal} {Science}\ }\textbf {\bibinfo {volume} {315}},\ \bibinfo {pages}
  {1116 } (\bibinfo {year} {2007})}\BibitemShut {NoStop}%
\bibitem [{\citenamefont {Kim}\ \emph {et~al.}(2012)\citenamefont {Kim},
  \citenamefont {Hanna}, \citenamefont {Byun}, \citenamefont {Santangelo},\
  and\ \citenamefont {Hayward}}]{Kim-Hanna}%
  \BibitemOpen
  \bibfield  {author} {\bibinfo {author} {\bibfnamefont {J.}~\bibnamefont
  {Kim}}, \bibinfo {author} {\bibfnamefont {J.}~\bibnamefont {Hanna}}, \bibinfo
  {author} {\bibfnamefont {M.}~\bibnamefont {Byun}}, \bibinfo {author}
  {\bibfnamefont {C.}~\bibnamefont {Santangelo}},\ and\ \bibinfo {author}
  {\bibfnamefont {R.}~\bibnamefont {Hayward}},\ }\bibfield  {title} {\bibinfo
  {title} {Designing responsive buckled surfaces by halftone gel lithography},\
  }\href {https://doi.org/10.1126/science.1215309} {\bibfield  {journal}
  {\bibinfo  {journal} {Science (New York, N.Y.)}\ }\textbf {\bibinfo {volume}
  {335}},\ \bibinfo {pages} {1201} (\bibinfo {year} {2012})}\BibitemShut
  {NoStop}%
\bibitem [{\citenamefont {Hu}\ \emph {et~al.}(2012)\citenamefont {Hu},
  \citenamefont {Meng}, \citenamefont {Li},\ and\ \citenamefont {Ibekwe}}]{Hu}%
  \BibitemOpen
  \bibfield  {author} {\bibinfo {author} {\bibfnamefont {J.}~\bibnamefont
  {Hu}}, \bibinfo {author} {\bibfnamefont {H.}~\bibnamefont {Meng}}, \bibinfo
  {author} {\bibfnamefont {G.}~\bibnamefont {Li}},\ and\ \bibinfo {author}
  {\bibfnamefont {S.~I.}\ \bibnamefont {Ibekwe}},\ }\bibfield  {title}
  {\bibinfo {title} {A review of stimuli-responsive polymers for smart textile
  applications},\ }\href {https://doi.org/10.1088/0964-1726/21/5/053001}
  {\bibfield  {journal} {\bibinfo  {journal} {Smart Materials and Structures}\
  }\textbf {\bibinfo {volume} {21}},\ \bibinfo {pages} {053001} (\bibinfo
  {year} {2012})}\BibitemShut {NoStop}%
\bibitem [{\citenamefont {Santangelo}(2017)}]{Santangelo}%
  \BibitemOpen
  \bibfield  {author} {\bibinfo {author} {\bibfnamefont {C.~D.}\ \bibnamefont
  {Santangelo}},\ }\bibfield  {title} {\bibinfo {title} {Extreme mechanics:
  Self-folding origami},\ }\href
  {https://doi.org/10.1146/annurev-conmatphys-031016-025316} {\bibfield
  {journal} {\bibinfo  {journal} {Annual Review of Condensed Matter Physics}\
  }\textbf {\bibinfo {volume} {8}},\ \bibinfo {pages} {165} (\bibinfo {year}
  {2017})}\BibitemShut {NoStop}%
\bibitem [{\citenamefont {Si{\'e}fert}\ \emph {et~al.}(2019)\citenamefont
  {Si{\'e}fert}, \citenamefont {Reyssat}, \citenamefont {Bico},\ and\
  \citenamefont {Roman}}]{Siefert}%
  \BibitemOpen
  \bibfield  {author} {\bibinfo {author} {\bibfnamefont {E.}~\bibnamefont
  {Si{\'e}fert}}, \bibinfo {author} {\bibfnamefont {E.}~\bibnamefont
  {Reyssat}}, \bibinfo {author} {\bibfnamefont {J.}~\bibnamefont {Bico}},\ and\
  \bibinfo {author} {\bibfnamefont {B.}~\bibnamefont {Roman}},\ }\bibfield
  {title} {\bibinfo {title} {Bio-inspired pneumatic shape-morphing
  elastomers},\ }\href {https://doi.org/10.1038/s41563-018-0219-x} {\bibfield
  {journal} {\bibinfo  {journal} {Nature materials}\ }\textbf {\bibinfo
  {volume} {18}},\ \bibinfo {pages} {24—28} (\bibinfo {year}
  {2019})}\BibitemShut {NoStop}%
\bibitem [{\citenamefont {Warner}\ and\ \citenamefont
  {Terentjev}(2007)}]{warner-book}%
  \BibitemOpen
  \bibfield  {author} {\bibinfo {author} {\bibfnamefont {M.}~\bibnamefont
  {Warner}}\ and\ \bibinfo {author} {\bibfnamefont {E.}~\bibnamefont
  {Terentjev}},\ }\href@noop {} {\emph {\bibinfo {title} {Liquid Crystal
  Elastomers}}},\ International Series of Monographs on Physics\ (\bibinfo
  {publisher} {OUP Oxford},\ \bibinfo {year} {2007})\BibitemShut {NoStop}%
\bibitem [{\citenamefont {Aharoni}\ \emph {et~al.}(2014)\citenamefont
  {Aharoni}, \citenamefont {Sharon},\ and\ \citenamefont
  {Kupferman}}]{aharoni-sharon-kupferman}%
  \BibitemOpen
  \bibfield  {author} {\bibinfo {author} {\bibfnamefont {H.}~\bibnamefont
  {Aharoni}}, \bibinfo {author} {\bibfnamefont {E.}~\bibnamefont {Sharon}},\
  and\ \bibinfo {author} {\bibfnamefont {R.}~\bibnamefont {Kupferman}},\
  }\bibfield  {title} {\bibinfo {title} {Geometry of thin nematic elastomer
  sheets},\ }\href {https://doi.org/10.1103/PhysRevLett.113.257801} {\bibfield
  {journal} {\bibinfo  {journal} {Phys. Rev. Lett.}\ }\textbf {\bibinfo
  {volume} {113}},\ \bibinfo {pages} {257801} (\bibinfo {year}
  {2014})}\BibitemShut {NoStop}%
\bibitem [{\citenamefont {Mostajeran}(2015)}]{mostajeran1}%
  \BibitemOpen
  \bibfield  {author} {\bibinfo {author} {\bibfnamefont {C.}~\bibnamefont
  {Mostajeran}},\ }\bibfield  {title} {\bibinfo {title} {Curvature generation
  in nematic surfaces},\ }\href {https://doi.org/10.1103/PhysRevE.91.062405}
  {\bibfield  {journal} {\bibinfo  {journal} {Phys. Rev. E}\ }\textbf {\bibinfo
  {volume} {91}},\ \bibinfo {pages} {062405} (\bibinfo {year}
  {2015})}\BibitemShut {NoStop}%
\bibitem [{\citenamefont {Duffy}\ \emph {et~al.}(2021)\citenamefont {Duffy},
  \citenamefont {Cmok}, \citenamefont {Biggins}, \citenamefont {Krishna},
  \citenamefont {Modes}, \citenamefont {Abdelrahman}, \citenamefont {Javed},
  \citenamefont {Ware}, \citenamefont {Feng},\ and\ \citenamefont
  {Warner}}]{duffy2}%
  \BibitemOpen
  \bibfield  {author} {\bibinfo {author} {\bibfnamefont {D.}~\bibnamefont
  {Duffy}}, \bibinfo {author} {\bibfnamefont {L.}~\bibnamefont {Cmok}},
  \bibinfo {author} {\bibfnamefont {J.~S.}\ \bibnamefont {Biggins}}, \bibinfo
  {author} {\bibfnamefont {A.}~\bibnamefont {Krishna}}, \bibinfo {author}
  {\bibfnamefont {C.~D.}\ \bibnamefont {Modes}}, \bibinfo {author}
  {\bibfnamefont {M.~K.}\ \bibnamefont {Abdelrahman}}, \bibinfo {author}
  {\bibfnamefont {M.}~\bibnamefont {Javed}}, \bibinfo {author} {\bibfnamefont
  {T.~H.}\ \bibnamefont {Ware}}, \bibinfo {author} {\bibfnamefont
  {F.}~\bibnamefont {Feng}},\ and\ \bibinfo {author} {\bibfnamefont
  {M.}~\bibnamefont {Warner}},\ }\bibfield  {title} {\bibinfo {title} {Shape
  programming lines of concentrated gaussian curvature},\ }\href
  {https://doi.org/10.1063/5.0044158} {\bibfield  {journal} {\bibinfo
  {journal} {Journal of Applied Physics}\ }\textbf {\bibinfo {volume} {129}},\
  \bibinfo {pages} {224701} (\bibinfo {year} {2021})}\BibitemShut {NoStop}%
\bibitem [{\citenamefont {Modes}\ \emph {et~al.}(2011)\citenamefont {Modes},
  \citenamefont {Bhattacharya},\ and\ \citenamefont {Warner}}]{modes-warner}%
  \BibitemOpen
  \bibfield  {author} {\bibinfo {author} {\bibfnamefont {C.~D.}\ \bibnamefont
  {Modes}}, \bibinfo {author} {\bibfnamefont {K.}~\bibnamefont
  {Bhattacharya}},\ and\ \bibinfo {author} {\bibfnamefont {M.}~\bibnamefont
  {Warner}},\ }\bibfield  {title} {\bibinfo {title} {Gaussian curvature from
  flat elastica sheets},\ }\href {http://www.jstor.org/stable/29792779}
  {\bibfield  {journal} {\bibinfo  {journal} {Proceedings: Mathematical,
  Physical and Engineering Sciences}\ }\textbf {\bibinfo {volume} {467}},\
  \bibinfo {pages} {1121} (\bibinfo {year} {2011})}\BibitemShut {NoStop}%
\bibitem [{\citenamefont {Mostajeran}\ \emph {et~al.}(2016)\citenamefont
  {Mostajeran}, \citenamefont {Warner}, \citenamefont {Ware},\ and\
  \citenamefont {White}}]{mostajeran2}%
  \BibitemOpen
  \bibfield  {author} {\bibinfo {author} {\bibfnamefont {C.}~\bibnamefont
  {Mostajeran}}, \bibinfo {author} {\bibfnamefont {M.}~\bibnamefont {Warner}},
  \bibinfo {author} {\bibfnamefont {T.~H.}\ \bibnamefont {Ware}},\ and\
  \bibinfo {author} {\bibfnamefont {T.~J.}\ \bibnamefont {White}},\ }\bibfield
  {title} {\bibinfo {title} {Encoding gaussian curvature in glassy and
  elastomeric liquid crystal solids},\ }\href@noop {} {\bibfield  {journal}
  {\bibinfo  {journal} {Proc. Math. Phys. Eng. Sci.}\ }\textbf {\bibinfo
  {volume} {472}},\ \bibinfo {pages} {20160112} (\bibinfo {year}
  {2016})}\BibitemShut {NoStop}%
\bibitem [{\citenamefont {Griniasty}\ \emph {et~al.}(2019)\citenamefont
  {Griniasty}, \citenamefont {Aharoni},\ and\ \citenamefont
  {Efrati}}]{griniasty-aharoni-efrati}%
  \BibitemOpen
  \bibfield  {author} {\bibinfo {author} {\bibfnamefont {I.}~\bibnamefont
  {Griniasty}}, \bibinfo {author} {\bibfnamefont {H.}~\bibnamefont {Aharoni}},\
  and\ \bibinfo {author} {\bibfnamefont {E.}~\bibnamefont {Efrati}},\
  }\bibfield  {title} {\bibinfo {title} {Curved geometries from planar director
  fields: Solving the two-dimensional inverse problem},\ }\href
  {https://doi.org/10.1103/PhysRevLett.123.127801} {\bibfield  {journal}
  {\bibinfo  {journal} {Phys. Rev. Lett.}\ }\textbf {\bibinfo {volume} {123}},\
  \bibinfo {pages} {127801} (\bibinfo {year} {2019})}\BibitemShut {NoStop}%
\bibitem [{\citenamefont {DeTurck}\ and\ \citenamefont
  {Yang}(1984)}]{deturck-yang}%
  \BibitemOpen
  \bibfield  {author} {\bibinfo {author} {\bibfnamefont {D.~M.}\ \bibnamefont
  {DeTurck}}\ and\ \bibinfo {author} {\bibfnamefont {D.}~\bibnamefont {Yang}},\
  }\bibfield  {title} {\bibinfo {title} {{Existence of elastic deformations
  with prescribed principal strains and triply orthogonal systems}},\ }\href
  {https://doi.org/10.1215/S0012-7094-84-05114-7} {\bibfield  {journal}
  {\bibinfo  {journal} {Duke Mathematical Journal}\ }\textbf {\bibinfo {volume}
  {51}},\ \bibinfo {pages} {243 } (\bibinfo {year} {1984})}\BibitemShut
  {NoStop}%
\bibitem [{\citenamefont {Gevirtz}(1993)}]{gevirtz93}%
  \BibitemOpen
  \bibfield  {author} {\bibinfo {author} {\bibfnamefont {J.}~\bibnamefont
  {Gevirtz}},\ }\bibfield  {title} {\bibinfo {title} {A diagonal hyperbolic
  system for mappings with prescribed principal strains},\ }\href
  {https://doi.org/https://doi.org/10.1006/jmaa.1993.1222} {\bibfield
  {journal} {\bibinfo  {journal} {Journal of Mathematical Analysis and
  Applications}\ }\textbf {\bibinfo {volume} {176}},\ \bibinfo {pages} {390}
  (\bibinfo {year} {1993})}\BibitemShut {NoStop}%
\bibitem [{\citenamefont {Aharoni}\ \emph {et~al.}(2018)\citenamefont
  {Aharoni}, \citenamefont {Xia}, \citenamefont {Zhang}, \citenamefont
  {Kamien},\ and\ \citenamefont {Yang}}]{aharoni2018universal}%
  \BibitemOpen
  \bibfield  {author} {\bibinfo {author} {\bibfnamefont {H.}~\bibnamefont
  {Aharoni}}, \bibinfo {author} {\bibfnamefont {Y.}~\bibnamefont {Xia}},
  \bibinfo {author} {\bibfnamefont {X.}~\bibnamefont {Zhang}}, \bibinfo
  {author} {\bibfnamefont {R.~D.}\ \bibnamefont {Kamien}},\ and\ \bibinfo
  {author} {\bibfnamefont {S.}~\bibnamefont {Yang}},\ }\bibfield  {title}
  {\bibinfo {title} {Universal inverse design of surfaces with thin nematic
  elastomer sheets},\ }\href {https://doi.org/10.1073/PNAS.1804702115}
  {\bibfield  {journal} {\bibinfo  {journal} {Proceedings of the National
  Academy of Sciences}\ }\textbf {\bibinfo {volume} {115}},\ \bibinfo {pages}
  {7206} (\bibinfo {year} {2018})}\BibitemShut {NoStop}%
\bibitem [{\citenamefont {Gevirtz}(1992)}]{gevirtz92}%
  \BibitemOpen
  \bibfield  {author} {\bibinfo {author} {\bibfnamefont {J.}~\bibnamefont
  {Gevirtz}},\ }\bibfield  {title} {\bibinfo {title} {On planar mappings with
  prescribed principal strains},\ }\href {https://doi.org/10.1007/BF00376186}
  {\bibfield  {journal} {\bibinfo  {journal} {Archive for Rational Mechanics
  and Analysis}\ }\textbf {\bibinfo {volume} {117}},\ \bibinfo {pages} {295}
  (\bibinfo {year} {1992})}\BibitemShut {NoStop}%
\bibitem [{\citenamefont {Gevirtz}(2002)}]{gevirtz02}%
  \BibitemOpen
  \bibfield  {author} {\bibinfo {author} {\bibfnamefont {J.}~\bibnamefont
  {Gevirtz}},\ }\bibfield  {title} {\bibinfo {title} {Boundary values and the
  transformation problem for constant principal strain mappings},\ }\href
  {https://doi.org/10.1155/S0161171203204166} {\bibfield  {journal} {\bibinfo
  {journal} {International Journal of Mathematics and Mathematical Sciences}\
  }\textbf {\bibinfo {volume} {2003}},\ \bibinfo {pages} {657924} (\bibinfo
  {year} {2002})}\BibitemShut {NoStop}%
\bibitem [{\citenamefont {Gevirtz}(2001)}]{gevirtz01}%
  \BibitemOpen
  \bibfield  {author} {\bibinfo {author} {\bibfnamefont {J.}~\bibnamefont
  {Gevirtz}},\ }\bibfield  {title} {\bibinfo {title} {Singularity sets of
  constant principal strain deformations},\ }\href
  {https://doi.org/https://doi.org/10.1006/jmaa.2001.7639} {\bibfield
  {journal} {\bibinfo  {journal} {Journal of Mathematical Analysis and
  Applications}\ }\textbf {\bibinfo {volume} {263}},\ \bibinfo {pages} {600}
  (\bibinfo {year} {2001})}\BibitemShut {NoStop}%
\bibitem [{\citenamefont {Gevirtz}(2008)}]{gevirtz08}%
  \BibitemOpen
  \bibfield  {author} {\bibinfo {author} {\bibfnamefont {J.}~\bibnamefont
  {Gevirtz}},\ }\bibfield  {title} {\bibinfo {title} {Boundary behavior of
  solutions of a class of genuinely nonlinear hyperbolic systems},\ }\href
  {https://doi.org/10.1137/070705507} {\bibfield  {journal} {\bibinfo
  {journal} {SIAM Journal on Mathematical Analysis}\ }\textbf {\bibinfo
  {volume} {40}},\ \bibinfo {pages} {1291} (\bibinfo {year}
  {2008})}\BibitemShut {NoStop}%
\bibitem [{\citenamefont {Niv}\ and\ \citenamefont
  {Efrati}(2018)}]{niv-efrati}%
  \BibitemOpen
  \bibfield  {author} {\bibinfo {author} {\bibfnamefont {I.}~\bibnamefont
  {Niv}}\ and\ \bibinfo {author} {\bibfnamefont {E.}~\bibnamefont {Efrati}},\
  }\bibfield  {title} {\bibinfo {title} {Geometric frustration and
  compatibility conditions for two-dimensional director fields},\ }\href
  {https://doi.org/10.1039/C7SM01672G} {\bibfield  {journal} {\bibinfo
  {journal} {Soft Matter}\ }\textbf {\bibinfo {volume} {14}},\ \bibinfo {pages}
  {424} (\bibinfo {year} {2018})}\BibitemShut {NoStop}%
\bibitem [{\citenamefont {Collins}(1982)}]{Collins}%
  \BibitemOpen
  \bibfield  {author} {\bibinfo {author} {\bibfnamefont {I.}~\bibnamefont
  {Collins}},\ }\bibfield  {title} {\bibinfo {title} {Boundary value problems
  in plane strain plasticity},\ }in\ \href@noop {} {\emph {\bibinfo {booktitle}
  {Mechanics of Solids}}},\ \bibinfo {editor} {edited by\ \bibinfo {editor}
  {\bibfnamefont {H.}~\bibnamefont {Hopkins}}\ and\ \bibinfo {editor}
  {\bibfnamefont {M.}~\bibnamefont {Swewll}}}\ (\bibinfo  {publisher}
  {Pergamon},\ \bibinfo {address} {Oxford},\ \bibinfo {year} {1982})\ pp.\
  \bibinfo {pages} {135--184}\BibitemShut {NoStop}%
\bibitem [{\citenamefont {Johnson}\ \emph {et~al.}(1970)\citenamefont
  {Johnson}, \citenamefont {Sowerby},\ and\ \citenamefont
  {Haddow}}]{johnson-book}%
  \BibitemOpen
  \bibfield  {author} {\bibinfo {author} {\bibfnamefont {W.}~\bibnamefont
  {Johnson}}, \bibinfo {author} {\bibfnamefont {R.}~\bibnamefont {Sowerby}},\
  and\ \bibinfo {author} {\bibfnamefont {J.}~\bibnamefont {Haddow}},\
  }\href@noop {} {\emph {\bibinfo {title} {Plane-strain Slip-line Fields: The
  Theory and Bibliography}}}\ (\bibinfo  {publisher} {American Elsevier
  Publishing Company},\ \bibinfo {year} {1970})\BibitemShut {NoStop}%
\bibitem [{\citenamefont {Graczykowski}\ and\ \citenamefont
  {Lewi{\'{n}}ski}(2006)}]{Graczykowski}%
  \BibitemOpen
  \bibfield  {author} {\bibinfo {author} {\bibfnamefont {C.}~\bibnamefont
  {Graczykowski}}\ and\ \bibinfo {author} {\bibfnamefont {T.}~\bibnamefont
  {Lewi{\'{n}}ski}},\ }\bibfield  {title} {\bibinfo {title} {Michell
  cantilevers constructed within trapezoidal domains-part {I}: Geometry of
  {H}encky nets},\ }\href@noop {} {\bibfield  {journal} {\bibinfo  {journal}
  {Structural and Multidisciplinary Optimization}\ }\textbf {\bibinfo {volume}
  {32}},\ \bibinfo {pages} {347} (\bibinfo {year} {2006})}\BibitemShut
  {NoStop}%
\bibitem [{\citenamefont {Acharya}(2019)}]{acharya}%
  \BibitemOpen
  \bibfield  {author} {\bibinfo {author} {\bibfnamefont {A.}~\bibnamefont
  {Acharya}},\ }\bibfield  {title} {\bibinfo {title} {A design principle for
  actuation of nematic glass sheets},\ }\href
  {https://doi.org/10.1007/s10659-018-9696-z} {\bibfield  {journal} {\bibinfo
  {journal} {Journal of Elasticity}\ }\textbf {\bibinfo {volume} {136}},\
  \bibinfo {pages} {237} (\bibinfo {year} {2019})}\BibitemShut {NoStop}%
\bibitem [{\citenamefont {Hencky}(1923)}]{Hencky}%
  \BibitemOpen
  \bibfield  {author} {\bibinfo {author} {\bibfnamefont {H.}~\bibnamefont
  {Hencky}},\ }\bibfield  {title} {\bibinfo {title} {{\"U}ber einige statisch
  bestimmte f{\"a}lle des gleichgewichts in plastischen k{\"o}rpern},\
  }\href@noop {} {\bibfield  {journal} {\bibinfo  {journal} {Zamm-zeitschrift
  Fur Angewandte Mathematik Und Mechanik}\ }\textbf {\bibinfo {volume} {3}},\
  \bibinfo {pages} {241} (\bibinfo {year} {1923})}\BibitemShut {NoStop}%
\bibitem [{\citenamefont {{Prandtl}}(1921)}]{Prandtl}%
  \BibitemOpen
  \bibfield  {author} {\bibinfo {author} {\bibfnamefont {L.}~\bibnamefont
  {{Prandtl}}},\ }\bibfield  {title} {\bibinfo {title} {{Hauptaufs{\"a}tze:
  {\"U}ber die Eindringungsfestigkeit (H{\"a}rte) plastischer Baustoffe und die
  Festigkeit von Schneiden}},\ }\href
  {https://doi.org/10.1002/zamm.19210010102} {\bibfield  {journal} {\bibinfo
  {journal} {Zeitschrift Angewandte Mathematik und Mechanik}\ }\textbf
  {\bibinfo {volume} {1}},\ \bibinfo {pages} {15} (\bibinfo {year}
  {1921})}\BibitemShut {NoStop}%
\bibitem [{\citenamefont {Michell}(1904)}]{Michell}%
  \BibitemOpen
  \bibfield  {author} {\bibinfo {author} {\bibfnamefont {A.}~\bibnamefont
  {Michell}},\ }\bibfield  {title} {\bibinfo {title} {The limits of economy of
  material in frame-structures},\ }\href
  {https://doi.org/10.1080/14786440409463229} {\bibfield  {journal} {\bibinfo
  {journal} {The London, Edinburgh, and Dublin Philosophical Magazine and
  Journal of Science}\ }\textbf {\bibinfo {volume} {8}},\ \bibinfo {pages}
  {589} (\bibinfo {year} {1904})}\BibitemShut {NoStop}%
\bibitem [{\citenamefont {Whittle}(2007)}]{whittle}%
  \BibitemOpen
  \bibfield  {author} {\bibinfo {author} {\bibfnamefont {P.}~\bibnamefont
  {Whittle}},\ }\href@noop {} {\emph {\bibinfo {title} {Networks: Optimisation
  and Evolution}}},\ Cambridge Series in Statistical and Probabilistic
  Mathematics\ (\bibinfo  {publisher} {Cambridge University Press},\ \bibinfo
  {year} {2007})\ pp.\ \bibinfo {pages} {95--115}\BibitemShut {NoStop}%
\bibitem [{\citenamefont {Lewi{\'n}ski}\ \emph {et~al.}(2018)\citenamefont
  {Lewi{\'n}ski}, \citenamefont {Sok{\'o}{\l}},\ and\ \citenamefont
  {Graczykowski}}]{michell-book}%
  \BibitemOpen
  \bibfield  {author} {\bibinfo {author} {\bibfnamefont {T.}~\bibnamefont
  {Lewi{\'n}ski}}, \bibinfo {author} {\bibfnamefont {T.}~\bibnamefont
  {Sok{\'o}{\l}}},\ and\ \bibinfo {author} {\bibfnamefont {C.}~\bibnamefont
  {Graczykowski}},\ }\href@noop {} {\emph {\bibinfo {title} {Michell
  Structures}}}\ (\bibinfo  {publisher} {Springer International Publishing},\
  \bibinfo {year} {2018})\BibitemShut {NoStop}%
\bibitem [{\citenamefont {Hill}(1998)}]{hill-book}%
  \BibitemOpen
  \bibfield  {author} {\bibinfo {author} {\bibfnamefont {R.}~\bibnamefont
  {Hill}},\ }\href@noop {} {\emph {\bibinfo {title} {The Mathematical Theory of
  Plasticity}}},\ Oxford classic texts in the physical sciences\ (\bibinfo
  {publisher} {Clarendon Press},\ \bibinfo {year} {1998})\ Chap.~\bibinfo
  {chapter} {6}\BibitemShut {NoStop}%
\bibitem [{\citenamefont {Collins}(1971)}]{Collins-plate}%
  \BibitemOpen
  \bibfield  {author} {\bibinfo {author} {\bibfnamefont {I.}~\bibnamefont
  {Collins}},\ }\bibfield  {title} {\bibinfo {title} {On an analogy between
  plane strain and plate bending solutions in rigid/perfect plasticity
  theory},\ }\href@noop {} {\bibfield  {journal} {\bibinfo  {journal}
  {International Journal of Solids and Structures}\ }\textbf {\bibinfo {volume}
  {7}},\ \bibinfo {pages} {1057} (\bibinfo {year} {1971})}\BibitemShut
  {NoStop}%
\bibitem [{\citenamefont {Geiringer}(1937)}]{Geiringer}%
  \BibitemOpen
  \bibfield  {author} {\bibinfo {author} {\bibfnamefont {H.}~\bibnamefont
  {Geiringer}},\ }\bibfield  {title} {\bibinfo {title} {Fondements
  math{\'e}matiques de la th{\'e}orie des corps plastiques isotropes},\
  }\href@noop {} {\bibfield  {journal} {\bibinfo  {journal} {M{\'e}morial des
  {S}ciences {M}ath{\'e}matiques,}\ }\textbf {\bibinfo {volume} {86}} (\bibinfo
  {year} {1937})}\BibitemShut {NoStop}%
\bibitem [{\citenamefont {Arcisz}\ and\ \citenamefont
  {Desperat}(1967)}]{Arcisz}%
  \BibitemOpen
  \bibfield  {author} {\bibinfo {author} {\bibfnamefont {M.}~\bibnamefont
  {Arcisz}}\ and\ \bibinfo {author} {\bibfnamefont {T.~W.}\ \bibnamefont
  {Desperat}},\ }\bibfield  {title} {\bibinfo {title} {On the
  {H}encky-{P}randtl nets based on two orthogonal circles},\ }\href
  {https://doi.org/10.1007/BF01178568} {\bibfield  {journal} {\bibinfo
  {journal} {Acta Mechanica}\ }\textbf {\bibinfo {volume} {4}},\ \bibinfo
  {pages} {205} (\bibinfo {year} {1967})}\BibitemShut {NoStop}%
\bibitem [{\citenamefont {Thomason}(1978)}]{Thomason}%
  \BibitemOpen
  \bibfield  {author} {\bibinfo {author} {\bibfnamefont {P.~F.}\ \bibnamefont
  {Thomason}},\ }\bibfield  {title} {\bibinfo {title} {{Riemann-Integral
  Solutions for the Plastic Slip-Line Fields Around Elliptical Holes}},\ }\href
  {https://doi.org/10.1115/1.3424381} {\bibfield  {journal} {\bibinfo
  {journal} {Journal of Applied Mechanics}\ }\textbf {\bibinfo {volume} {45}},\
  \bibinfo {pages} {678} (\bibinfo {year} {1978})}\BibitemShut {NoStop}%
\end{thebibliography}%
